\documentclass[letterpaper,11pt]{article}
\usepackage[margin=1in]{geometry}
\usepackage[utf8]{inputenc} 
\usepackage[T1]{fontenc}    
\usepackage[colorlinks=true,breaklinks=true,bookmarks=true,urlcolor=blue,citecolor=blue,linkcolor=blue,bookmarksopen=false,draft=false]{hyperref}
\usepackage{url}            
\usepackage{booktabs}       
\usepackage{multirow}
\usepackage{microtype}      
\usepackage{amsfonts,amsmath, amsthm,mathtools,stmaryrd}
\usepackage{xcolor}
\usepackage[inline]{enumitem}
\usepackage{appendix}
\usepackage{float, graphicx, subcaption}
\usepackage{bbold}
\usepackage{natbib}
\usepackage{bbm}
\renewcommand{\cite}[1]{\citep{#1}}

\newtheorem{remark}{Remark}[section]

\newtheorem{proposition}{Proposition}[section]

\newtheorem{definition}{Definition}[section]

\def\argmin_#1{\underset{#1}{\mathrm{arg\,min\, }}}
\def\argmax_#1{\underset{#1}{\mathrm{arg\,max\, }}}

\makeatletter
\def\dasharrowfill@#1#2#3#4{%
        $\m@th
        \thickmuskip0mu
        \medmuskip\thickmuskip
        \thinmuskip\thickmuskip
        \relax
        #4#1\mkern2mu
        \xleaders\hbox{$#4\mkern2mu#2\mkern2mu$}\hfill
        \mkern2mu
        #3$%
}

\def\dashleftarrowfill@{\dasharrowfill@\leftarrow\relbar\relbar}
\def\dashrightarrowfill@{\dasharrowfill@\relbar\relbar\rightarrow}
\def\dashleftrightarrowfill@{\dasharrowfill@\leftarrow\relbar\rightarrow}
\def\dashLeftarrowfill@{\dasharrowfill@\Leftarrow\Relbar\Relbar}
\def\dashRightarrowfill@{\dasharrowfill@\Relbar\Relbar\Rightarrow}
\def\dashLeftrightarrowfill@{\dasharrowfill@\Leftarrow\Relbar\Rightarrow}

\providecommand*\xdashleftarrow[2][]{%
  \ext@arrow 0055{\dashleftarrowfill@}{#1}{#2}}
\providecommand*\xdashrightarrow[2][]{%
  \ext@arrow 0055{\dashrightarrowfill@}{#1}{#2}}
\providecommand*\xdashleftrightarrow[2][]{%
  \ext@arrow 0055{\dashleftrightarrowfill@}{#1}{#2}}
\providecommand*\xdashLeftarrow[2][]{%
  \ext@arrow 0055{\dashLeftarrowfill@}{#1}{#2}}
\providecommand*\xdashRightarrow[2][]{%
  \ext@arrow 0055{\dashRightarrowfill@}{#1}{#2}}
\providecommand*\xdashLeftrightarrow[2][]{%
  \ext@arrow 0055{\dashLeftrightarrowfill@}{#1}{#2}}
\makeatother
\usepackage{nicefrac}       
\usepackage{wrapfig}         
\usepackage{algorithm}
\usepackage{algpseudocode}
\usepackage{caption}
\usepackage{mwe}
\begin{document}
\title{Generation of Threat: Crediting football players for creating dangerous actions in
an unbiased way}

\author{
Ali Baouan
\thanks{Centre de Math\'ematiques Appliqu\'ees, Ecole Polytechnique.
\textbf{Email:}
ali.baouan@polytechnique.edu}
\and
Sebastien Coustou
\thanks{Parma Calcio 1913 Performance and Analytics.
\textbf{Email:} 
scoustou@parmacalcio1913.com }
\and
Mathieu Lacome
\thanks{Parma Calcio 1913 Performance and Analytics.
\textbf{Email:} 
mlacome@parmacalcio1913.com }
\and 
Sergio Pulido
\thanks{Université Paris–Saclay, CNRS, ENSIIE, Univ Évry, LaMME.
\textbf{Email:} 
sergio.pulidonino@ensiie.fr }
\and
Mathieu Rosenbaum
\thanks{Centre de Math\'ematiques Appliqu\'ees, Ecole Polytechnique.
\textbf{Email:}
mathieu.rosenbaum@polytechnique.edu}
}
\date{}
\maketitle

\begin{abstract}
We introduce an innovative methodology to identify football players at the origin of threatening actions in a team. In our framework, a threat is defined as entering the opposing team's \textit{danger area}. We investigate the timing of threat events and ball touches of players, and capture their correlation using Hawkes processes. Our model-based approach allows us to evaluate a player's ability to create danger both directly and through interactions with teammates. We define a new index, called \textit{Generation of Threat} (GoT), that measures in an unbiased way the contribution of a player to threat generation. For illustration, we present a detailed analysis of Chelsea's 2016-2017 season, with a standout performance from Eden Hazard. We are able to credit each player for his involvement in  danger creation and determine the main circuits leading to threat. In the same spirit, we investigate the danger generation process of Stade Rennais in the 2021-2022 season. Furthermore, we establish a comprehensive ranking of Ligue 1 players based on their generated threat in the 2021-2022 season. Our analysis reveals surprising results, with players such as Jason Berthomier, Moses Simon and Frederic Guilbert among the top performers in the GoT rankings.
\end{abstract}
\section{Introduction}
Which player should be credited for a successful action or sequence in a football match? In the case of a goal, the striker obviously plays an important role. However, we all have in mind goals where the striker just needs to push the ball after a great assist. In that case, the passer is certainly the most important player involved. Some argue that the second-to-last pass is actually the most crucial component as it is often this pass that creates  disequilibrium. Sometimes, we even see a clearance by a goalkeeper being at the origin of a dangerous situation.
\\
\noindent
\\
In this work, our goal is to build a quantitative and unbiased methodology enabling us to assess the importance of a player in the generation of dangerous actions. By a threat, we simply mean a situation where a player of the team of interest gets the ball in the danger area of the opposing team. The danger area is defined as a rectangular region around the opponent's goal where the likelihood of scoring from a shot is high. To achieve our objective, we need to model interactions between players, taking into account past events in the game accurately. This is because we want, for example, to be able to credit a defender for a great pass that leads to a dangerous situation after several ball touches following the initial pass. Therefore, at the timestamp where the action is considered dangerous (in our case when the ball reaches the danger area), we must "remember" the original pass of the defender.
\\
\noindent
\\
Thus, at a given time $t$, we want to draw links between past events in the game and its future. With this objective in mind, simply relying on the current state of the game (players and ball's positions) as the information set is not enough for modeling the game accurately.  It is important to consider the dynamics that occurred prior to time $t$.  This is in contrast to the so-called Markovian approach where one summarizes information obtained from the beginning of the game until time $t$ by the state of the game at time $t$. The Markovian setting is in fact underlying some very relevant and successful metrics introduced recently such as the expected goals \cite{xgblog} and expected assists \cite{xablog}. For example, the expected goal estimates the probability that a shot results in a goal based on factors such as the distance to the goal and the angle of the shot, both attributes of the game state at time $t$. The Markov assumption is in that case natural as these features give a reasonable estimate of the quality of the chance.
Similarly, the expected assists aim at measuring the probability that a pass leads to a goal, by looking at a different subset of game state features, such as the type of the pass and the coordinates of the target. What these two approaches have in common is that given time $t$ they define a value for an action (pass or shot), that is determined by the game state at time $t$ only and does not look at the past patterns of play. In the same spirit, the expected threat introduced in \cite{xtblog} assigns a value to each game state depending only on the position of the ball. This value combines the possibilities of a direct shot or a pass to another position in quantifying the expected number of goals. 
\\
\noindent
\\
To account for the effect of past events in the future dynamics of a game, we introduce Hawkes processes \cite{hawkes1971point,hawkes1971spectra} to reproduce interactions between players. Hawkes processes are stochastic models used to model sequences of random events. They are widely used in various fields such as earthquake modeling \cite{adamopoulos1976cluster,ogata1988statistical}, neuroscience \cite{lambert2018reconstructing,bonnet2022neuronal} and finance  \cite{jaisson2015limit}. In our case, the events are the times when players touch the ball. Specifically, we implement a Hawkes process with 11 components (number of players in the team), with component $i$ corresponding to player $i$ of the team of interest. The value of this component at time $t$ is simply the number of times player $i$ has touched the ball from the beginning of the game to time $t$. At each time the player touches the ball, his corresponding component increases by one. The innovation here is that we collect information from these timestamps and their correlations from one player to the other teammates.
\\
\noindent
\\
The specificity of Hawkes processes is that at time $t$, the probability that player $i$ gets the ball shortly after $t$ depends on which players had possession of the ball before $t$ and how long ago they had it. The impact on this probability of a player touching the ball a long time before $t$ is negligible compared to a player who had possession right before $t$. The ability to reproduce the decaying impact of events with time is a particularly useful property of Hawkes processes in our context. For instance, let us consider a central defender. At time $t$, the probability that he gets the ball in the near future should be high if, in the recent past (last few seconds), he already touched the ball and/or another central defender did. On the contrary, if the forward players have held the ball for the past minute, this probability should be low.
\\
\noindent
\\
Then, we add a twelfth component to our Hawkes process that we call threat. The value of the threat component at time $t$ is simply the number of times the ball has reached the danger area of the opposing team between the beginning of the game and time $t$. Treating this component as part of our Hawkes process, we are able to model the influence of each player in the generation of threat.
\\
\noindent
\\
Calibrating our model allows us to assess the contribution of each player of a team to the creation of dangerous situations. We are therefore able to investigate carefully the subtle dynamics and connections leading to ominous situations. In particular, we can emphasize the crucial role of certain players that are not spotted by other statistics. Note that our calibration requires the analysis of a data set of at least ten games. So we are not evaluating each action occurring in a game but rather the global performance of players in terms of threat generation over a sequence of games.
\\
\noindent
\\
 More precisely, the structure of Hawkes processes allows us to define the Generation of Threat (GoT) indices to objectively evaluate a player's involvement in the creation of  threats over a considered series of games. These metrics quantify the expected number of dangerous situations for which a player can be credited. The direct generation of threat indices GoT$^{(dir)}$ and GoT$^{(dir)}_{90}$ measure the number of threats the player is directly responsible for generating per touch of the ball and per 90 minutes, respectively. Directly generating a threat can be viewed as being the last link in the chain of events leading to it. On the other hand, the indirect generation of threat indices GoT$^{(ind)}$ and GoT$^{(ind)}_{90}$ measure the indirect contribution per touch and per 90 minutes, respectively, adding the danger created via the interactions with other players too. In this case, we count all the instances where the player participates in the chain of events leading to the dangerous situation. As an application, we use the GoT indices to rank the Ligue 1 players in the 2021-2022 season. Not surprisingly, the top positions are dominated by established offensive players. However, we also identify some surprising picks, including Jason Berthomier, Moses Simon and Frederic Guilbert, who rank among the top twenty players. We also compare the performance of the Ligue 1 central defenders in terms of GoT$^{(ind)}_{90}$. Naturally, defenders from Paris Saint-Germain stand out and benefit from the offensive performance of their forwards. However, we also identify other excellent center-back pairs such as  Nayef Aguerd and Warmed Omari from Stade Rennais, and Facundo Medina and Jonathan Gradit from Lens. Moreover, our approach allows us to rate these players based on their performance in specific positions in a formation, providing a tool to identify the optimal position for each player.
 \\
\noindent
\\
 Our approach has the property of being easily interpretable using the immigration-birth representation of linear Hawkes processes, see \cite{hawkes1974cluster}.  This representation induces a notion of hierarchy between events and allows us to visualize the interactions between different event types in a graph. All player touches can be viewed as individuals in a population, and each individual independently generates offsprings, that are threat events or ball touches of the same player or other players. In particular, this enables us to effectively interpret the estimated GoT metrics as a measure of the hierarchical relationship between the player's touch and subsequent threat events. Furthermore, we can construct interaction networks of football teams and graphically analyze a team's in-game dynamics and danger creation circuits. We apply this approach to investigate games from Chelsea in the 2016-2017 season and Stade Rennais in the 2021-2022 season. We are able to effectively capture the main threat creation circuits that the opponent should try to control. Identifying specific patterns and evaluating the ability of players to create threat with our methodology paves the way to more informed decisions about tactics. 
\\
\noindent
\\
The article is organized as follows. In Section \ref{sec:data}, we describe the event-based data we have in hand and how it is processed. We also provide an overview of Hawkes processes and recall the results that are useful for our football application. Furthermore, we present the interpretation of the estimated parameters in the context of football and define the Generation of Threat (GoT) metrics. In Section \ref{sec:estimation}, we briefly describe the maximum likelihood estimation methodology. We also conduct a study on simulated data to measure estimation accuracy that can be expected on real datasets depending on the amount of available data. We find that reliable estimation can be obtained from  600 minutes of football data. Section \ref{sec:chelsea} presents the results of our analysis on a collection of Chelsea games in the 2016-2017 season. In Section \ref{sec:ligue1}, we establish a ranking of Ligue 1 players in the 2021-2022 season based on their GoT indices. Finally, in the appendix, we present the analysis of the Stade Rennais games in the 2021-2022 Ligue 1 season.
\section{Modeling event-based football data}
\label{sec:data}
This section provides a short overview of Hawkes processes as applied to football event data. It includes necessary definitions and theoretical results for a better understanding of the subsequent analysis of football dynamics. Furthermore, we discuss the interpretation of the parameters of the Hawkes process in this context and introduce the Generation of Threat (GoT) metrics to quantify a player’s influence on creating dangerous situations.
\subsection{Description of the data}
We use the  F24 files provided by Stats-perform. Each file gives comprehensive information about a football match, including the formation of each team and the position of each player on the pitch. Unlike real-time tracking datasets, the F24 files do not record the continuous movement of all players; instead, they focus on discrete events. Specifically, this dataset is a time-coded feed that lists all player action events within the game, detailing the player involved, team affiliation, event type, the coordinates on the pitch and the timestamp. Event types include passes, shots, goals, set pieces, etc. Each event is further described by a series of qualifiers that provide additional context and specifics. 
\\
\noindent
\\
The temporal nature of the dataset naturally encourages the use of point process models to analyze in-game events. In the Stats-perform classification system, each position on the pitch is assigned a number $p$ in  $\{1,\dots,11\}$ for each formation. Therefore, we construct a multivariate counting process that tracks the number of events associated with the player in each position $p$ at all times $t$. Analyzing the relationships between the components of this process can provide interesting insights into the playing dynamics of the team. To capture the impact of the ball touches of each player on the offensive performance, we use Hawkes processes to model the constructed counting processes. Details on how the counting process is constructed are provided in Section~\ref{sec:processing}.

\subsection{Hawkes processes}
\label{sec:hawkes}
As mentioned in the introduction, Hawkes processes are a class of multivariate point processes introduced in \cite{hawkes1971point}. If we consider a vector  $N(t)=(N_i(t))_{i\in \{1,\dots,d\}}$, where $N_i(t)$ denotes the number of events for the $i$-th component between $0$ and $t$, the associated intensity process can essentially be defined as: 
\begin{equation*}
\lambda_i\left(t \right):=\lim _{h \rightarrow 0^+} \frac{ \mathbb{P}( N_i(t+h)-N_i(t)=1 | \mathcal{F}_{t})}{h} .
\end{equation*}
\noindent
Here,  $\mathcal{F}_t$ is the filtration generated by $\{N_s , \ s<t\}$, that is the information set available at time $t$. The intensity of a counting process determines the rate at which new jumps occur based on past events, see \cite{bremaud1981point} for a more rigorous definition. In the case of Hawkes processes, the intensity is a linear combination of past jump times. 
\begin{definition}[Hawkes process]
A $d$-variate Hawkes process is a counting-process $N(t) \in \mathbb{R}^d$ whose $i$-th component is determined by its intensity of the form:
\begin{equation*}
    \lambda_i (t)=\mu_i + \sum_{j=1}^d \sum_{t_k^{(j)}<t} \phi_{i,j}(t-t_k^{(j)}) ,
\end{equation*}
where the $\left ( t_k^{(j)}\right )_{k \geq 1}$ are the times of events for dimension $j$ for $j=1,\dots,d$. $\mu_i \in \mathbb{R}^+$  is a constant baseline intensity and $\phi_{i,j} : \mathbb{R}^+  \rightarrow \mathbb{R}^+$ is a non-negative kernel. We can write the expression for the intensity in the vectorial form:
\begin{equation*}
    \lambda(t)=\mu + \int_0^t \phi(t-s) dN(s),
\end{equation*}
with $\mu \in \mathbb{R}^{+d}$  and $\phi=\left \{ \phi_{i,j} \right\}_{0\leq i,j\leq d} : \mathbb{R}^+ \rightarrow \mathbb{R}^{d\times d} $ a non-negative matrix-valued kernel.
\end{definition}
\noindent
 The underlying idea behind Hawkes processes is that a constant intensity $\mu$ generates the initial batch of jumps across all dimensions. These jumps are random but the rate of their occurrence remains constant over time. Then, each jump increases the intensity in the near future; therefore, exciting new jumps, that in turn trigger other jumps. This leads to a chain reaction called the self-excitation property of Hawkes processes.
 \\
 \noindent
 \\
In this work, the counting process $N(t)$ records the ball touches of  each player and the self-excitation property is desirable.  In fact, whenever a player touches the ball, it raises the probability of further touches—either by the same player or by teammates—in the near future. Consequently, we expect diagonal components $\phi_{i,i}$ of the kernel to be significant, since a player’s own touches often occur in short succession. More generally, the off-diagonal kernels $\phi_{i,j}$ reflect how a touch from player $j$ increases the likelihood of a subsequent touch from player $i$. To quantify the level of interaction between the components of a Hawkes process, we introduce the notion of branching matrix. This definition is also necessary to state conditions for the system to be stable.
\begin{definition}[Branching matrix, stability]
\label{def:branching}
The branching matrix of a Hawkes process is defined as,
\begin{equation*}
    K=\int_0^{\infty} \phi(t)dt =\left \{ \int_0^\infty \phi_{i,j}(t) dt\right \}_{1 \leq i,j\leq d}.
\end{equation*}
Moreover, a Hawkes process is said to be stable if $\int_0^\infty \phi_{i,j}(t) dt<\infty$ for all $i,j$ and if the spectral radius $\rho(K)$ of the branching matrix satisfies: \begin{equation*}
    \rho(K)<1.
\end{equation*}
See \cite{jaisson2015limit} for more details.
\end{definition}
\noindent
In the context of football event-based data, $K$ provides us with a measure of the interaction between players.  The higher the value of $K_{i,j}$, the more likely there are ball touches from player $i$ following a ball touch from player $j$. 
Finally, we use the parametric class of exponential kernels in our estimation methodology. 
\begin{definition}[Exponential kernels]
The exponential kernel is defined as 
\begin{equation*}
    \phi_{i,j}(t)=\alpha_{i,j} e^{-\beta_{i,j}t} 1_{t\geq 0},
\end{equation*}
where $\alpha_{i,j}$, $\beta_{i,j}$ are nonnegative real numbers.
\end{definition}
\noindent
Exponential kernels are particularly nice from a computational viewpoint in estimation. Additionally, their parameters are easy to interpret. In fact, the branching matrix in this case is simply given by $K=(\frac{\alpha_{i,j}}{\beta_{i,j}})_{i,j}$ and the decay parameter $\beta_{i,j}$ indicates the speed at which cross excitation decreases.
\subsection{Processing of the data for Hawkes inference}
\label{sec:processing}
\begin{figure}[t!]
    \centering
    \begin{minipage}{0.9\textwidth}
        \centering
        \scalebox{0.6}{\includegraphics{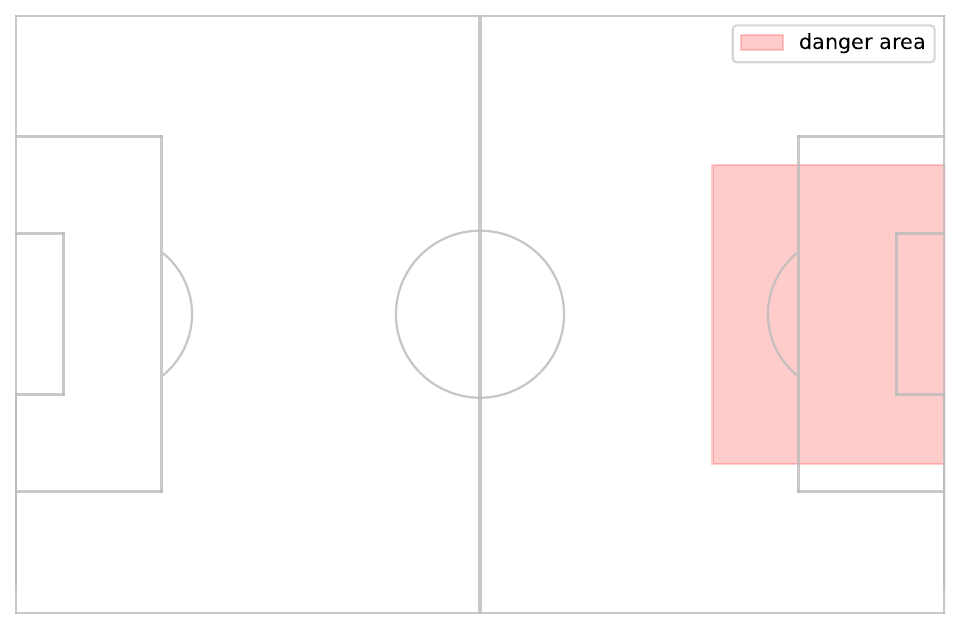}}\vspace*{-.5em}
        \captionof{figure}{Representation of the danger area.}\label{fig:dangerarea}
    \end{minipage}
\end{figure}

\begin{figure}[t!]
    \centering
    \begin{minipage}{0.9\textwidth}
        \centering
        \scalebox{0.46}{\includegraphics{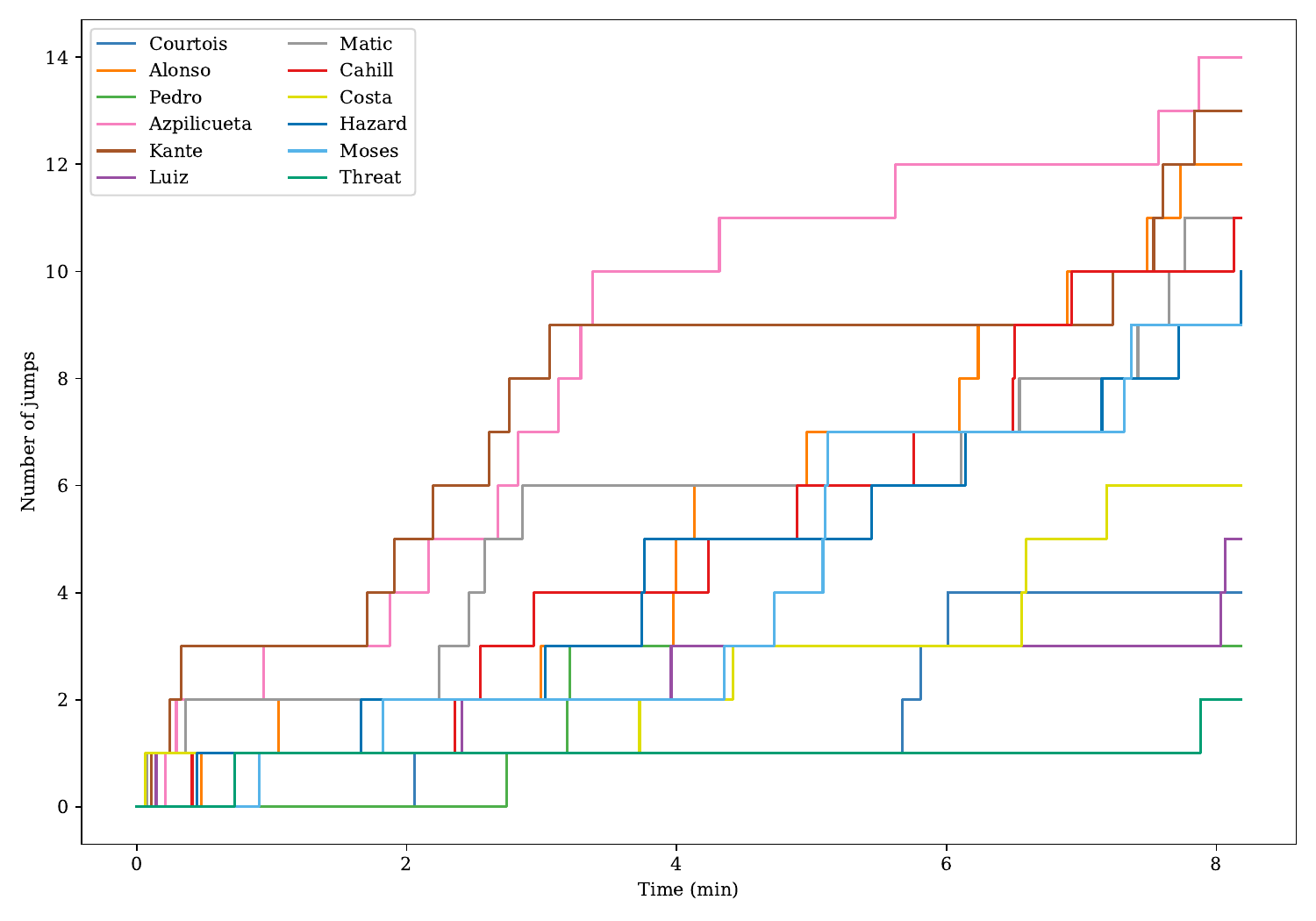}}\vspace*{-.5em}
        \captionof{figure}{Staircase plots illustrating the cumulative number of events for each of the $11$ players and the threat event over the first $8$ minutes of the game of Chelsea against Manchester United on October 2016.}\label{fig:examplepointprocess}
    \end{minipage}
\end{figure}

We study our event-based data  using Hawkes processes. Doing so, we can gain insights from timestamps of events and information about the spatial coordinates of the ball. For each game, the players occupying each position are provided, with each position assigned a number $p$ in  $\{1,\dots,11\}$ following the Stats-perform classification system. For a given team and a list of its games, we build a 12-dimensional point process for each game. Each dimension $p\in \{1,\dots,11\}$ records the timestamps of ball touches by the player occupying position $p$, regardless of his identity. The twelfth dimension represents the threat state and is triggered every time there is a ball touch by a player from the considered team in the danger area of the opponent.  The danger area is defined as a box around the opposing goal covering 50\% of the width of the pitch and 25\% of its length, as illustrated in Figure \ref{fig:dangerarea}. When a player has possession of the ball in this region, the probability of a shot occurring is high, see \cite{xtblog}  for an estimate of the shot probability at each location on the pitch. Compared to the penalty surface, the danger area is slightly closer to the midfielders and defenders, enabling us to capture more threat events generated by these positions.
\\
\noindent
\\
The following rules are applied when constructing the process:
\begin{enumerate}
    \item Every time a player in the considered team touches the ball, there is a jump in the dimension $p\in\{1,\dots,11\}$ associated with his position.
    \item Every time a player in the considered team touches the ball inside the opposing threat area, there is a jump in the twelfth dimension at the corresponding timestamp. In this case, no jump is recorded in the component associated with the player. 
    \item \label{enum:1} Once a threat state is triggered, no jumps or time are recorded until the ball exits the danger area. We resume counting the jumps when the ball is outside the danger area by at least two meters.
    \item \label{enum:4} When the ball is lost (when there is an event where the opposing team has the ball), the time and events are not recorded until the ball is won again. Upon regaining possession, we resume recording the events in our point process by adding a random duration,  with an average of twelve seconds, generated from the sum of two exponential distributions of parameter six.
    \item \label{enum:5}  We exclude crossing events coming from a free kick or a corner.
\end{enumerate}
\noindent
 Rule \ref{enum:1} is considered to avoid consecutive threat states. We are not interested in the auto-exciting property of the threat events. Therefore, we stop recording once a threat state is achieved and only resume when the team is outside the opposing surface by at least two meters. In Rule \ref{enum:4}, we want to avoid having large durations where no event occurs. This is the case every time the considered team loses the ball to the opposition. Thus, the possession times of the opponent are compressed into an average of twelve seconds. The choice of the twelve seconds threshold  is based on the average duration between events to which we add another exponential random variable as a penalization for losing the ball. The constructed point process considers possession stretches of the team to be uninterrupted. Rule \ref{enum:5} is implemented because the crossing events are highly correlated with threat events. In particular, the designated set piece taker of each team is naturally responsible for more threats. Therefore, we choose to discard these events to remove bias from our measure of danger creation and ensure fair player comparisons. 
 \\
 \noindent
 \\
 Figure~\ref{fig:examplepointprocess} displays an example of the resulting point process in the game of Chelsea against Manchester United on October 2022. Each line tracks the cumulative number of ball touches of a player, with the final line corresponding to the threat event. Given a collection of games of a team, the point processes built from each game are assembled into one process.  The aim is to determine the hierarchical relationship between these player touches by examining the timing of their occurence. Additionally, we are interested in identifying the positions where a ball touch is directly correlated to a future jump in the twelfth dimension, which represents a threat. We also want to measure the indirect contribution of a player to the generation of threat through his interaction with other players.  

\begin{figure}[t!]
        \centering
        \begin{subfigure}[b]{0.475\textwidth}
            \centering
            \includegraphics[width=\textwidth]{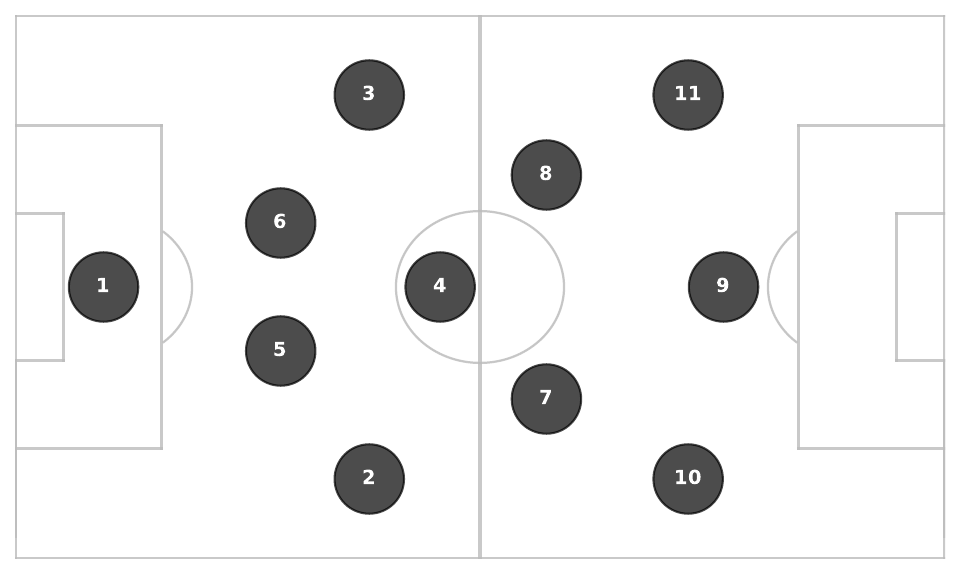}
            \caption[Network2]%
            {{\small 433 Formation.}}    
            \label{fig:433}
        \end{subfigure}
        \hfill
        \begin{subfigure}[b]{0.475\textwidth}  
            \centering 
            \includegraphics[width=\textwidth]{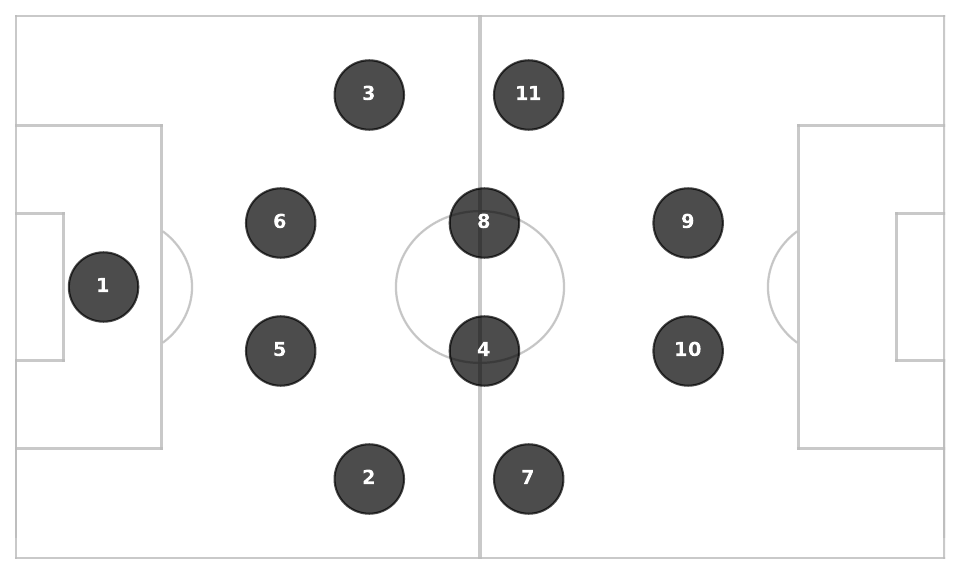}
            \caption[]%
            {{\small 442 Formation.}}    
            \label{fig:442}
        \end{subfigure}
        \vskip\baselineskip
        \begin{subfigure}[b]{0.475\textwidth}   
            \centering 
            \includegraphics[width=\textwidth]{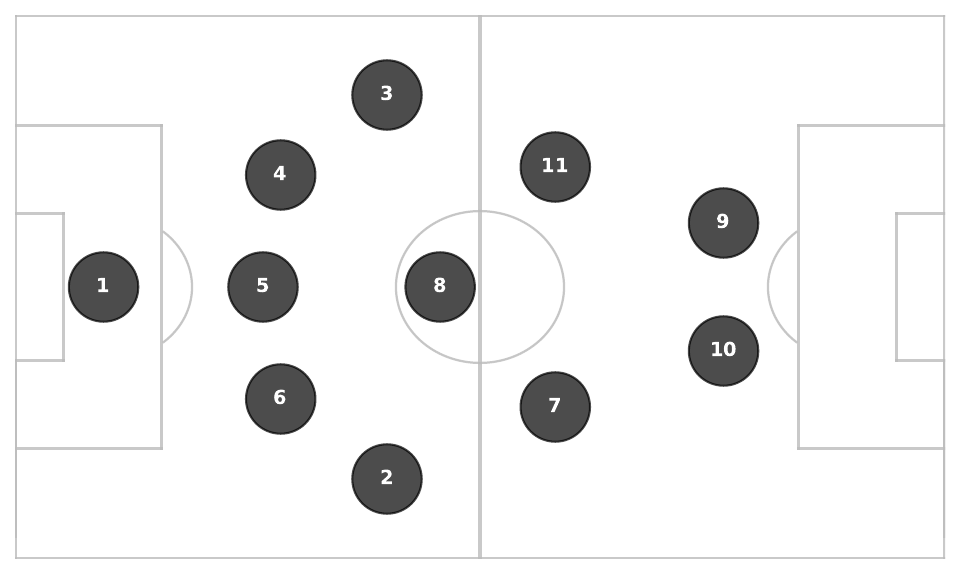}
            \caption{{\small 532 Formation.}}    
            \label{fig:532}
        \end{subfigure}
        \hfill
        \begin{subfigure}[b]{0.475\textwidth}   
            \centering 
            \includegraphics[width=\textwidth]{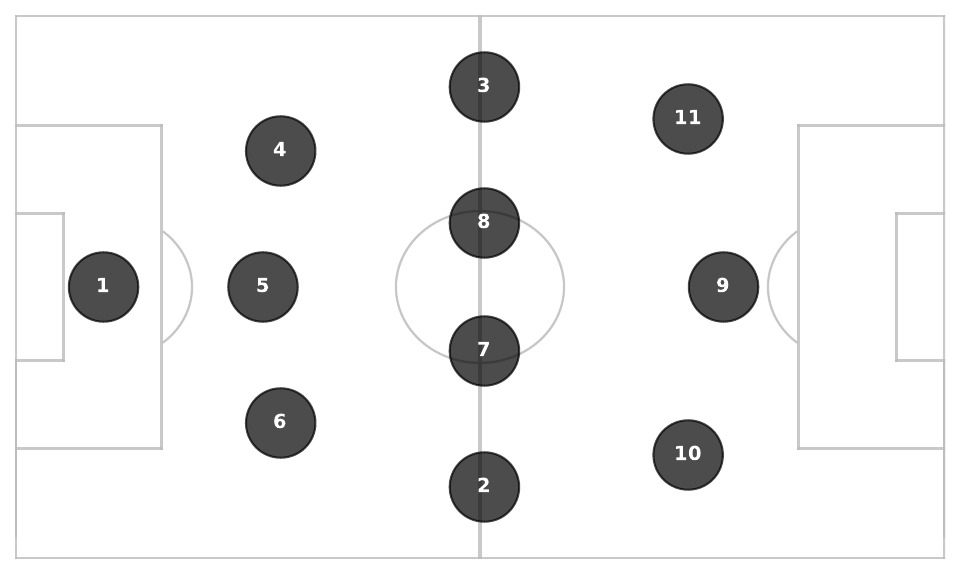}
            \caption{{\small 343 Formation.}}    
            \label{fig:343}
        \end{subfigure}
        \caption{\small The number associated with each position for each group of formations.} 
        \label{fig:formation_dimensions}
    \end{figure}

\begin{remark}[A different twelfth dimension]
In this work, we have incorporated a twelfth dimension that tracks the instances of entering the opposing danger area. This is done because we want to identify the players who are responsible for creating the threat events. Our approach can be extended for various analyses by selecting an alternative twelfth state. For example, we can choose to record the timestamps of ball losses in the twelfth dimension instead of threats. This would enable us to identify the players who are most accountable for losing possession and measure the correlation between their touches and subsequent turnovers.
\end{remark}
\noindent
Since we assemble point processes from different games, it is essential to ensure that the components of these processes correspond to the same positions across all games. In particular, the role associated with the position number $p$ can vary depending on the formation used. To maintain homogeneity between assembled game segments, the analysis is conducted only on games where each position represents the same role. This is determined by looking at the formation used in the game. Given the large number of formations and their similarity, we group them into clusters based on their structural shapes. Within each cluster, formations have the same number of players in each line, allowing for a one-to-one assignment between corresponding positions. Subsequently, we use matches from the most commonly used cluster for each team. The clusters are determined as follows: 
\begin{itemize}
    \item Cluster 1: 433, 4141, 4231, 4321.
    \item Cluster 2: 442, 41212, 451, 4411, 4222.
    \item Cluster 3: 532, 352, 31312, 3511, 3412.
    \item Cluster 4: 343, 541, 3421.
\end{itemize}
The distribution of the positions for each formation clusters is shown in Figure \ref{fig:formation_dimensions}.
\begin{remark}
In the following, we aim at evaluating a player's performance when he plays in a specific position. To achieve this, we only consider sequences of games where the player in question is playing in that position. We record ball touches in the other positions regardless of the identity of the player occupying them. In Section \ref{sec:chelsea} and Appendix \ref{app:rennes} where we analyze the interactions between the starting eleven players in given teams, we only record sequences of games where the same eleven players play in their respective positions.
\end{remark}

\subsection{Immigration-birth representation}
\label{sec:immigrant}
Introduced in \cite{hawkes1974cluster}, the immigration-birth representation provides an intuitive way to understand linear Hawkes processes. Let us consider a stable $d$-variate Hawkes process $N(t)$ with a baseline intensity $\mu$ and  a  kernel $\phi$.  The law of such point process can be described through a population approach. Essentially, we consider a population where immigrants of $d$ types arrive at random times. Each of them gives birth to children of all types. Then the children, grand-children, grand-grand-children etc. also give birth to children of all types.  More precisely, the dynamic is constructed as follows:
 \begin{itemize}
\item For $j=1,2,\dots,d$ , we consider an instance of a Poisson process with rate $\mu_j$, with its elements called immigrants of type $j$. Generation $0$ consists of the immigrants;

\item Recursively, given generations $0,1,\dots,n$, each individual born at time $s$ of type $j$  in generation $n$ generates its offspring of type $i$ as an independent instance of a non-homogeneous Poisson process with rate $\lambda_t^{s,n}:=\phi_{i,j}(t-s)$ for $t\geq s$. The union of these offspring of all types constitutes generation $n+1$.

\item The point process is then defined as the union of all generations.
\end{itemize}
\noindent
The resulting process has the law of  a Hawkes process. In this representation, stability means each individual has less than one child on average in the case $d=1$, which ensures some good mathematical properties for the process. From now on, we assume that all considered Hawkes processes are stable. Additionally, under this construction, $K_{i,j}=\int_0^\infty \phi_{i,j}(t)dt$ can be interpreted as the expected number of direct children of type $i$ of an individual of type $j$. The following proposition provides a closed-form formula for the expected number of descendants of a single individual. It includes both immediate descendants and those from later generations. This result is derived similarly to the one-dimensional case in \cite{jaisson2015limit}, and allows us to quantify the average number of events originating from each jump from each dimension.
\begin{proposition}
\label{prop:totalnum}
The entry $i,j$ of the matrix $K(I-K)^{-1}$ gives  the expected number of descendants of type $i$ generated by an individual of type $j$. 
\end{proposition}
\noindent
The immigration-birth representation is particularly powerful in the context of football event modeling as it captures the hierarchical and interdependent nature of in-game actions. Each event can be viewed as an individual that potentially triggers a cascade of subsequent events. This framework allows us to model not only the direct influence of a player's action on future events but also the indirect effects by looking at its descendants.

\subsection{Generation of Threat (GoT) indices}
The immigration-birth representation of Hawkes processes explained in Section \ref{sec:immigrant} allows us to establish connections between the events in a football match. Essentially, each ball touch or threat event can be seen as an individual in a population, that generates first-generation children of various types - ball touches from other players and threat situations. These offspring, in turn, generate additional ball touches or threat events etc. When we say that an event generates a ball touch or a dangerous situation, we mean that it is responsible for its occurrence. This is a subtle definition because being responsible for an action does not necessarily mean providing the pass that leads to it. In some instances, the second-to-last pass is the most crucial step in creating the dangerous situation. There may even be several events between the generating ball touch and the dangerous action. Our approach eliminates these "noisy" in-between events and associates events through parent-child connections. Hawkes processes impute the responsibility of generating a threat to the most likely parent event, even if it occurred prior to other ball touches. In particular, they allow us to quantify the average number of dangerous actions that can be attributed to a given player. 
\\
\noindent
\\Using this population representation, we define the following GoT indices to assess the  ability of a player to generate threat when he plays in a given position. The first two indices evaluate the impact of one touch of the player whereas the latter two measure the impact of the player's touches over 90 minutes.

\paragraph{Direct GoT per touch (GoT$^{(dir)}$):} A ball touch from the player in position $p$ generates first-generation children of type threat. We refer to these instances as the \textit{direct} threat events generated by the player touch. We define GoT$^{(dir)}$ as the average number of these threat events that occur because of one touch from player $p$. This metric describes the intrinsic ability of the player to create dangerous situations. It can be calculated through the estimated branching matrix:
\begin{equation*}
    \textup{GoT}^{(dir)}(p)= K_{12,p}.
\end{equation*}

\paragraph{Indirect GoT per touch (GoT$^{(ind)}$):} A ball touch from a given player can be directly responsible for a threat event, but can also generate other ball touches that then generate danger. To quantify the total impact of a single player touch on the danger creation process, we use Proposition \ref{prop:totalnum} and consider the matrix $$M=K(I-K)^{-1}.$$
 \noindent
The coefficient $M_{12, p}$ represents the expected number of threat events where the ball touch from the player $p$ originates the chain of events leading to it. This includes the threat directly generated but also the one resulting from a sequence of other player touches. The difference with the GoT$^{(dir)}$ index is that we credit the player touch for being at the root of the generation process and not for the crucial creative step.
\begin{equation*}
    \textup{GoT}^{(ind)}(p)= M_{12,p}.
\end{equation*}
\paragraph{Direct GoT per 90 minutes (GoT$^{(dir)}_{90}$):} We may want to account for the involvement in the game of a given player by normalizing GoT$^{(dir)}$ by his expected number of touches. We define the direct GoT per 90 minutes as the expected number of dangerous actions over 90 minutes\footnote{Note that here 90 minutes corresponds to 90 minutes of data after processing which does not translate to 90 minutes in a football match. This is notably because of the concatenation of sequences of possession explained in Section \ref{sec:processing}.} for which we credit the player:
\begin{equation*}
   \textup{GoT}_{90}^{(dir)}= \mathbf{E}\left (N_{p}(T) \right ) \times \textup{GoT}^{(dir)}(p),
\end{equation*}
where $T=90\textup{ minutes}$. The expected number of touches vector can be approximated thanks to the law of large numbers: 
\begin{equation*}
    \mathbf{E}\left (N(T) \right ) \approx (I-K)^{-1} \mu T.
\end{equation*}

\paragraph{Indirect GoT per 90 minutes (GoT$^{(ind)}_{90}$):} This index measures the expected number of threats over 90 minutes where a given player is involved in the building circuit. We define the indirect GoT per 90 minutes as the average number of threat events subtracted by the average number of threat events if the considered player is removed from the pitch. The GoT$^{(ind)}_{90}$ index is therefore calculated as follows: 
\begin{equation*}
\textup{GoT}_{90}^{(ind)}=\mathbf{E}\left ( N_{12}(T,K,\mu)-N_{12}(T,K^{(-p)},\mu^{(-p)}) \right ),
\end{equation*}
where $K^{(-p)}$ is defined as the matrix K where the $p^{th}$ row and $p^{th}$ column are set to zero. Likewise, $\mu^{(-p)}$ is defined as the vector $\mu$ where the $p^{th}$ coordinate is set to 0. The expected number of threats can be approximated using the branching matrix and the baseline intensity $\mu$:
\begin{equation*}
    \mathbf{E}\left (N_{12}(T,K,\mu) \right )\approx \left ((I-K)^{-1} \mu T \right ) _{12}.
\end{equation*}

\begin{remark}
Calculating the GoT$^{(ind)}_{90}$ by multiplying the GoT$^{(ind)}$ index by the average number of ball touches of the player would overestimate the player's involvement in danger creation. In fact, we would count multiple times the circuits leading to threat where the player touches the ball more than once.
\end{remark}
\noindent
Additionally, a ball touch from a player can also be responsible for generating ball touches from other players or himself. In this case as well, this is not necessarily achieved through a direct pass. Hawkes processes allow us to estimate the expected number of these generated ball touches.  Similar to the  GoT$^{(dir)}$ index definition,  the branching coefficient $K_{p_1,p_2}$ indicates the expected number of touches of player $p_1$ that happen because a given ball touch from player $p_2$ occurred before. The graphical representation of these interaction indices through a graph helps us gain a better understanding of the danger creation process. In particular, it allows us to identify the patterns of play that end in a threat.

\section{Maximum Likelihood estimation}\label{sec:estimation}
\subsection{Likelihood of Hawkes process}

This section describes briefly parameters estimation for multivariate Hawkes processes, see \cite{ogata1978estimators,bonnet2022inference}. Consider a $d$-variate point process $(N(t))$ on $[0,T]$ with intensity of the form
\begin{equation*}
    \lambda_i (t,\theta^*)=\mu_i^* + \sum_{j=1}^d \sum_{t_k^{(j)}<t} \alpha_{i,j}^* \exp \left (-\beta_{i,j}^*(t-t_k^{(j)} ) \right )  ,
\end{equation*}\noindent
where $\theta^*=(\mu^*,\alpha^*,\beta^*)$ are some unknown parameters. Given fixed parameters $\theta=(\mu,\alpha,\beta)$ and a realization of the Hawkes process, the log-likelihood is calculated as follows:
\begin{equation}
\label{eq:likelihood}
    \ell(\theta) = \sum_{i=1}^d \left (-\int_0^T \lambda_i(s,\theta)ds +\sum_{t_k^{(j)}<T} \log\left (\lambda_i(t_k^{(i)},\theta) \right)   \right ).
\end{equation}
The maximum likelihood estimator is the parameter that maximizes the above function. It can be observed from Equation (\ref{eq:likelihood}) that the likelihood can be separated into $d$ distinct subfunctions, each dependent on the parameters $\mu_i$ and $(\alpha_{i,j},\beta_{i,j})_{j=1,\dots,d}$ for $i$ in $\{1,\dots,d\}$. As a result, the optimization can be performed separately $d$ different times to estimate each subset of parameters. It is shown in \cite{ogata1978estimators} that this estimator is consistent. Additionally, the log-likelihood can be simplified in the case of exponential kernels and computed in time complexity of $\mathcal{O} \left (  d^2 N(T)  \right )$, see \cite{ogata1981lewis}. For example, for $d=1$ and $T=t_n$, the likelihood is given by :

\begin{equation*}
     \ell(\theta) = \sum\limits_{i=1}^{n} \log \left ( \mu + \alpha R(i) \right ) -\mu t_n + \frac{\alpha}{\beta} \sum_{i=1}^n \left ( e^{-\beta (t_n-t_i)} - 1 \right),
\end{equation*}
where $R(i)= \sum\limits_{j=1}^{i-1} e^{-\beta (t_i-t_j)}$ can be computed recursively for i in $\{2,\dots,n\}$ :

\begin{equation*}
     R(i)=e^{-\beta(t_i-t_{i-1})} \left ( 1+R(i-1) \right ).
\end{equation*}

\begin{remark}
The likelihood function is not concave with respect to $(\beta_{k,l})_{k,l=1,\dots,d}$ in the exponential case. This means that convergence to the global maximum is not guaranteed, especially in large dimensions. Fixing $\beta_{k,l}=\beta_{k}$ for all $l=1,\dots,d$ as proposed by \cite{bonnet2022inference} produces very good results for $d=12$. In this case, each of the objective functions is not concave in only one parameter instead of $d$.
\end{remark}

\begin{remark}
In the context of football, the effect of a ball touch on the intensity of the process should last no longer than a few seconds. When $n$ realizations of football matches are concatenated and treated as one long game, the likelihood function should not be altered by much. In fact, the rapid decay of the exponential kernel compared to the duration of games makes the induced error negligible. 
\end{remark}

\subsection{Simulation study}\label{subsec:toy}
\begin{figure}[t!]
    \centering
    \begin{subfigure}[b]{0.45\textwidth}
        \centering
        \includegraphics[width=\textwidth]{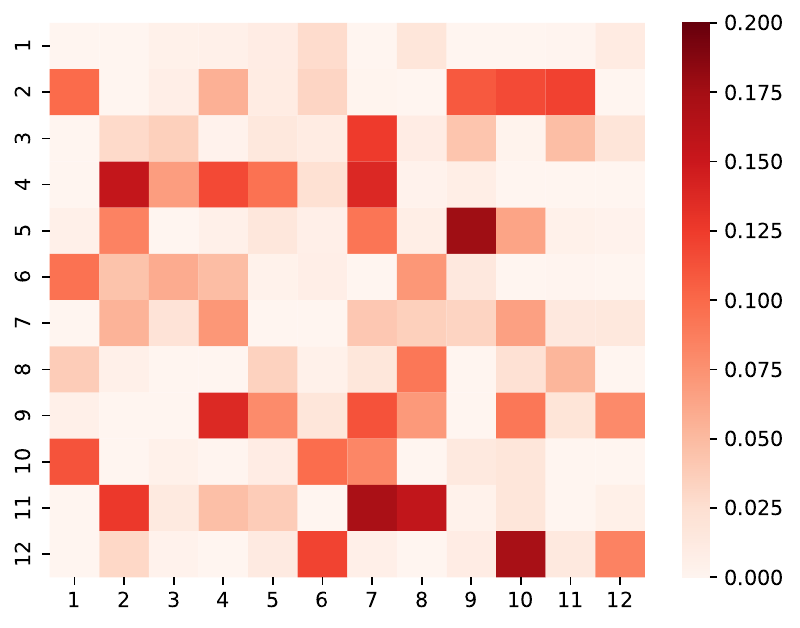}
        \caption{Estimated branching matrix.}
        \label{fig:testestim}
    \end{subfigure}
    \hfill
    \begin{subfigure}[b]{0.45\textwidth}
        \centering
        \includegraphics[width=\textwidth]{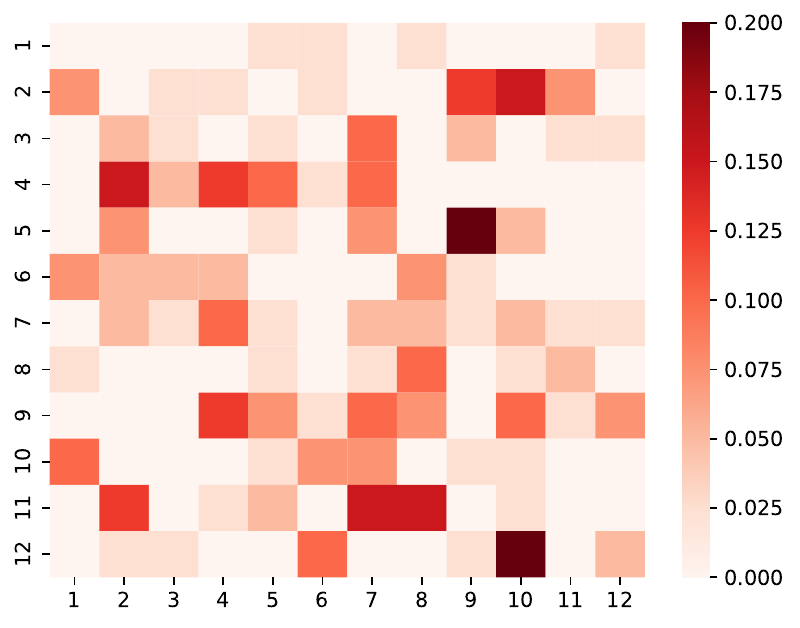}
        \caption{True branching matrix.}
        \label{fig:testtrue}
    \end{subfigure}
    \caption{Heatmaps of the the true branching matrix and the estimated one from a simulation of horizon $600$ minutes.}
    \label{fig:gaussian_comparison}
\end{figure}

The goal of this section is to evaluate the maximum likelihood estimation using a simulated dataset that reproduces similar dynamics  as those in a football game. We want to determine the amount of data required for an accurate estimation of the branching matrix. We also want to assess the model's ability to detect a null kernel between two dimensions. A null kernel $\phi_{i,j}$ means a jump in dimension $j$ has no exciting effect on dimension $i$. In the context of football, it is particularly informative to detect such an absence of connection between players. 
 \\
 \noindent
 \\
We perform  $100$ simulations over different horizons. For each simulation, the parameters are sampled as follows: 
\begin{itemize}
    \item $\mu$ is chosen from a uniform random variable over $[0.006,0.01]$.
    \item $\beta$ is chosen to be constant for all $i,j$ in   $\{1,\dots,12\}$ sampled from a uniform random variable over $[0.5,1]$ .
    \item The $\alpha_{i,j}$ are chosen independently from a geometric distribution of parameter $p=0.4$ scaled by $40 $ for all $i,j$ so that $40\% $ of the values are equal to $0$. 
\end{itemize}
\begin{table}[!ht]
    \centering
    \begin{tabular}{c|ccc}
    \toprule
    \toprule
        \textbf{Horizon (minutes)} & \textbf{False positive} & \textbf{Error on false negative} & \textbf{Relative error} \\ \midrule
        \bfseries 300 & 3.1\% & 0.0094 & 27.0\% \\ 
        \midrule%
        \bfseries 600 & 0.4\% & 0.0063 & 18.6\% \\
        \midrule
        \bfseries 1200 &  0.0\% & 0.0043 &  13.4\% \\ 
        \midrule%
        \bfseries 2400 &  0.0\% & 0.0030 & 9.5\% \\
        \bottomrule
        \bottomrule
    \end{tabular}
    \vspace{0.5em}
    \caption{Accuracy results of the maximum likelihood estimation of Hawkes parameters on the simulated datasets.}
    \label{tab:testresults}
\end{table}
Then we fit a 12-dimensional Hawkes process to this data using the algorithm from \cite{bonnet2022inference}. We analyze the resulting accuracy as a function of the simulation horizon. Table \ref{tab:testresults} presents the results through three different metrics, averaged accross the $100$ simulations for each horizon value:
\begin{itemize}
    \item False positive: Percentage of branching matrix coefficients $\hat{K}_{i,j}=\frac{\alpha_{i,j}}{\beta_{i,j}}$ wrongly estimated as null when $K_{i,j}> 0$. The estimated parameter is considered null if its value is not statistically significant, with a threshold set at $10^{-8}$. Our estimation correctly detects existing links even for small horizons.
    \item Error on false negative: Our estimation consistently detects  at least 50\% of null links $K_{i,j}=0$. The estimation on the remaining 50\% is generally very low as can be seen in Table \ref{tab:testresults}.
    \item Relative error:  The weighted mean absolute percentage error when $K_{i,j}>0$. This metric is defined as the mean absolute error divided by the average value of $K_{i,j}$: 
    $$\textup{wMAPE}=\frac{\sum\limits_{i,j} \left|\hat{K}_{i,j} -K_{i,j} \right | \mathbbm{1} _{K_{i,j}>0} }{ \sum\limits_{i,j} K_{i,j} \mathbbm{1} _{K_{i,j}>0} }.$$
\end{itemize}
The maximum likelihood estimate is good enough for our purposes given the high dimensionality. Figure \ref{fig:testestim} shows the estimated branching matrix from a simulation of horizon 600 minutes. We observe that the estimated branching matrix appears to correctly approximate the true branching matrix in Figure \ref{fig:testtrue}.

\section{Analysis of Chelsea FC in the 2016-2017 season}
\label{sec:chelsea}
As a first example, we perform our analysis on a selection of Chelsea FC matches from the 2016-2017 season. The team had a stable formation and a constant starting eleven over thirteen games in the Premier League. This is quite convenient because we retrieve a large amount of data where each position $p$ in $\{1,\dots,11\}$ is associated with one player. Similar analysis for Stade Rennais in the 2021-2022 season is provided in Appendix \ref{app:rennes}.

\subsection{Selected games}
In Table \ref{tab:selectedchelsea}, we give the list of selected games for Chelsea FC. In each of these games, the flat 343 formation is used for at least sixty minutes and the starting eleven remains the same: 
\begin{itemize}
    \item Thibaut Courtois.
    \item Gary Cahill - David Luiz - Cesar Azpilicueta.
    \item Marcos Alonso - Nemanja Matic - N'Golo Kante - Victor Moses. 
    \item Eden Hazard - Diego Costa - Pedro Rodriguez.
\end{itemize}
\noindent
Therefore, we use the data before the first substitution from Chelsea FC in each game to build the counting process.

\begin{table}[t!]
    \centering
     \resizebox{0.6\columnwidth}{!}{\begin{tabular}{llll}
    \toprule
    \toprule
        \textbf{Date} & \textbf{Opponent} & \textbf{Home or Away} & \textbf{Competition} \\ 
        \midrule
        Oct 15, 2016 & Leicester City & Home & English Premier League \\ 
        Oct 23, 2016 & Manchester United & Home & English Premier League \\ 
        Oct 30, 2016 & Southampton & Away & English Premier League \\ 
        \midrule
        Nov 5, 2016 & Everton & Home & English Premier League \\ 
        Nov 20, 2016 & Middlesbrough & Away & English Premier League \\ 
        Nov 26, 2016 & Tottenham Hotspur & Home & English Premier League \\ 
        \midrule
        Dec 11, 2016 & West Bromwich Albion & Home & English Premier League \\ 
        Jan 4, 2017 & Tottenham Hotspur & Away & English Premier League \\ 
        Jan 22, 2017 & Hull City & Home & English Premier League \\ 
        \midrule
        Feb 4, 2017 & Arsenal & Home & English Premier League \\ 
        Feb 12, 2017 & Burnley & Away & English Premier League \\ 
        Apr 8, 2017 & Bournemouth & Away & English Premier League \\ 
        \midrule
        Apr 30, 2017 & Everton & Away & English Premier League \\
        \bottomrule
        \bottomrule
    \end{tabular}}
    \vspace{0.5em}
    \caption{List of selected games with the same starting eleven for Chelsea FC.}
    \label{tab:selectedchelsea}
\end{table}

\subsection{Results and discussion}

\begin{figure}[!ht]
    \centering
    \begin{minipage}{0.9\textwidth}
        \centering
        \scalebox{0.6}{\includegraphics{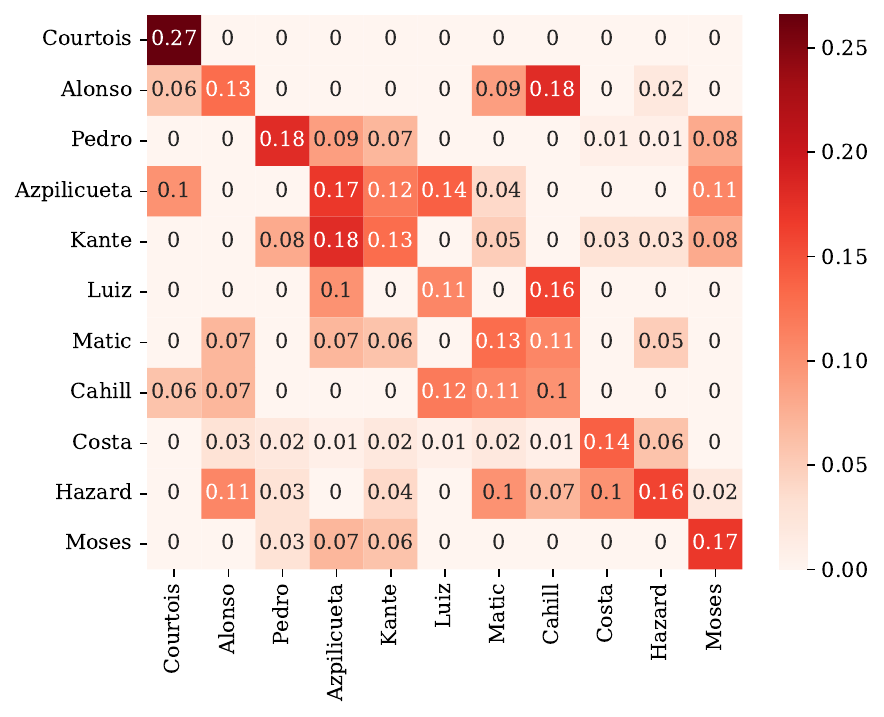}}\vspace*{-.5em}
        \captionof{figure}{Estimated branching matrix for Chelsea FC.}\label{fig:bmchelsea}      
    \end{minipage}
\end{figure}

In Table \ref{tab:dangerchelsea}, we display the different GoT indices for the Chelsea players. We include Monte Carlo estimates of the standard error (SE) for the indices. Using the estimated parameters, we generate $100$ sample paths of the same length as the original dataset, and then the parameters are re-estimated again. Figure \ref{fig:graphchelsea} graphically represents  the direct interactions between players as well as their GoT$^{(ind)}$ indices and Figure \ref{fig:bmchelsea} shows the estimated branching matrix. We can identify two buildup schemes along the wings with two triangles: Cahill-Alonso-Matic  and Kante-Azpilicueta-Moses. The main channel of communication between both sides is based on the Matic-Kante link.
\\
\noindent
\\
Below is a list of observations on players: 
\begin{figure}[t!]
    \centering
    \begin{minipage}{0.9\textwidth}
        \centering
        \scalebox{0.6}{\includegraphics{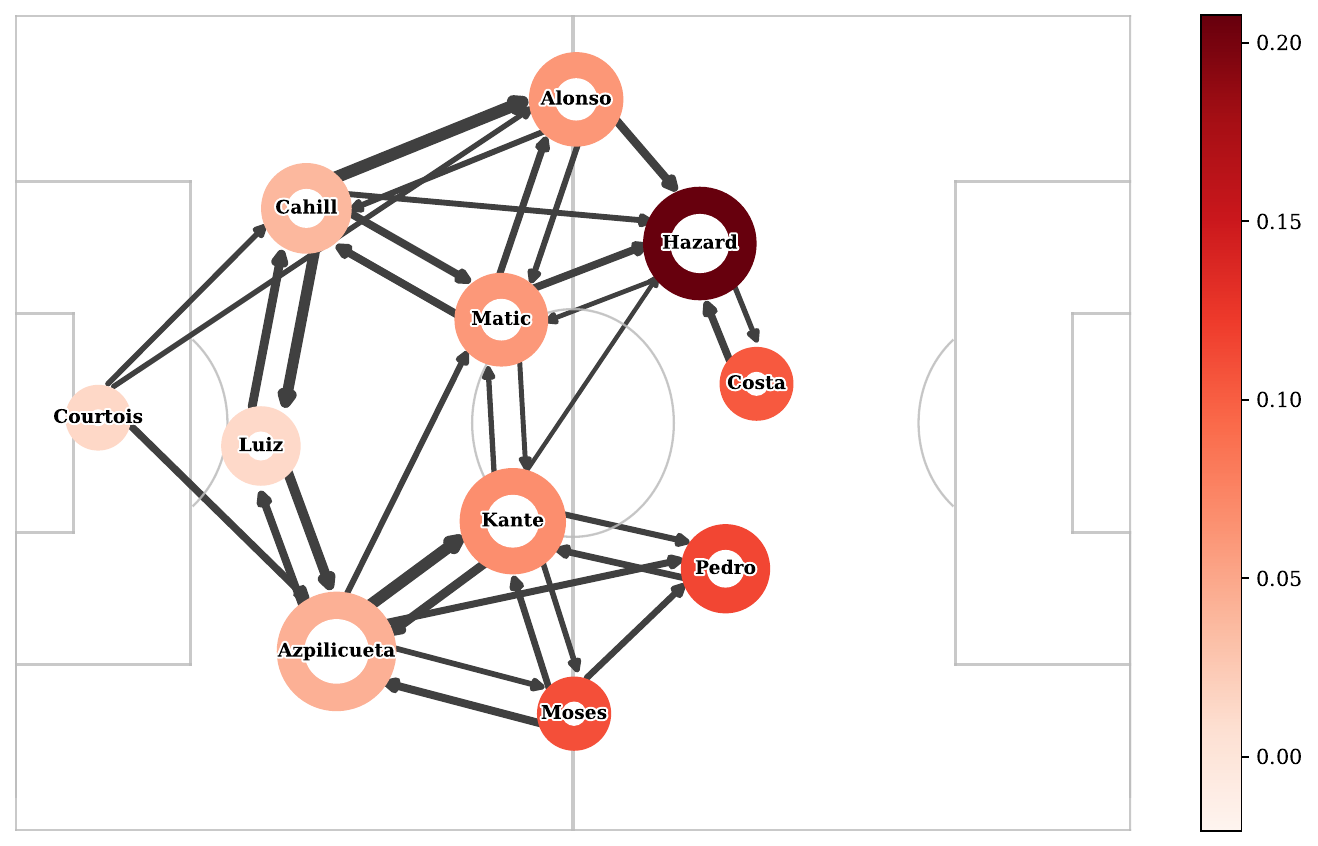}}\vspace*{-.5em}
        \captionof{figure}{Graph summarizing the interactions between Chelsea players. The width of an arrow from player $p_1$ to player $p_2$ is proportional to the expected number of touches of player $p_2$ generated by one touch from player $p_1$. The size of the circle of player $p$ is proportional to the sum of the arrow sizes received, indicating the involvement of the player in the considered games. The color of the circle represents the GoT$^{i}$ index for each player.}\label{fig:graphchelsea}
    \end{minipage}
\end{figure}
\begin{table}[!ht]
    \centering
    \resizebox{0.6\columnwidth}{!}{\begin{tabular}{lllll}
\toprule
\toprule
\bfseries Player name & \bfseries GoT$^{(dir)}$  {\tiny (SE) }& \bfseries GoT$^{(ind)}$  {\tiny (SE) }& \bfseries GoT$^{(dir)}_{90}$  {\tiny (SE) }& \bfseries GoT$^{(ind)}_{90}$  {\tiny (SE) }\\
\midrule
\bfseries Eden Hazard & 0.16   {\tiny (0.020)} & 0.21   {\tiny (0.025)} & 14.2   {\tiny (2.019)} & 15.0   {\tiny (2.060)} \\
\bfseries Victor Moses & 0.07   {\tiny (0.017)} & 0.11   {\tiny (0.021)} & 5.7   {\tiny (1.471)} & 7.5   {\tiny (1.500)} \\
\bfseries Pedro Rodriguez & 0.08  {\tiny (0.016)} & 0.12   {\tiny (0.020)} & 5.5   {\tiny (1.218)} & 6.7   {\tiny (1.277)} \\
\midrule
\bfseries N'Golo Kante & 0.02   {\tiny (0.012)} & 0.07   {\tiny (0.016)} & 2.7   {\tiny (1.351)} & 6.2   {\tiny (1.399)} \\
\bfseries Nemanja Matic & 0.01  {\tiny (0.010)} & 0.06   {\tiny (0.014)} & 1.5   {\tiny (1.082)} & 5.2   {\tiny (1.244)} \\
\bfseries Marcos Alonso & 0.02   {\tiny (0.011)} & 0.06   {\tiny (0.016)} & 1.9   {\tiny (1.095)} & 5.1   {\tiny (1.296)} \\
\midrule
\bfseries Diego Costa & 0.07   {\tiny (0.022)} & 0.10   {\tiny (0.028)} & 3.6   {\tiny (1.214)} 	& 4.8   {\tiny (1.256)} \\
\bfseries Cesar Azpilicueta & 0.00   {\tiny (0.005)} 	& 0.04   {\tiny (0.009)} & 0.0   {\tiny (0.581)} & 4.1   {\tiny (0.812)} \\
\bfseries Gary Cahill & 0.00  {\tiny (0.006)} & 0.04   {\tiny (0.010)} & 0.0   {\tiny (0.550)} & 3.0   {\tiny (0.842)} \\
\midrule
\bfseries David Luiz & 0.00   {\tiny (0.005)} & 0.01   {\tiny (0.008)} & 0.0   {\tiny (0.492)} & 1.0   {\tiny (0.591)} \\
\bfseries Thibaut Courtois & 0.00  {\tiny (0.006)} & 0.01   {\tiny (0.009)} & 0.0   {\tiny (0.334)} & 0.6   {\tiny (0.373)} \\
\bottomrule
\bottomrule
\end{tabular}}
\vspace{0.3em}
        \caption{Generated threat metrics for the players of Chelsea FC. The table is sorted by GoT$^{(ind)}_{90}$.}
    \label{tab:dangerchelsea}
\end{table}

\paragraph{Eden Hazard:} Unsurprisingly, the offensive player, ranked second in the PFA Players' Player of the Year 2017 award, leads all GoT metrics. In particular, there is no significant difference between his GoT$^{(dir)}_{90}$ and GoT$^{(ind)}_{90}$ indices, indicating that his primary way of creating danger is through direct threat. Hazard was well known for his aggressive and direct play as well as for his dribbling.

\paragraph{N'Golo Kante:} Ranking fourth in GoT$^{(ind)}_{90}$ is evidence to Kante's important role in Chelsea's success in the 2016-2017 season. The winner of the PFA Players' Player of the Year 2017 award is definitely not limited to defense as the numbers show that he is largely involved in danger creation. This is explained by the fact that Kante is a box to box midfielder and that he is at the center of multiple circuits that end in a threat:
\begin{itemize}
    \item Kante $\rightarrow$ Pedro $\rightarrow$ Threat.
    \item Kante $\rightarrow$ Moses $\rightarrow$ Pedro $\rightarrow$ Threat.
    \item Kante $\rightarrow$ Matic $\rightarrow$ Hazard $\rightarrow$ Threat.
\end{itemize}
\paragraph{David Luiz:} The contribution of the central defender David Luiz in the generation of threat is minimal. This is not surprising as the flat 3-4-3 system relies heavily on the wings. David Luiz naturally passes the ball to either Gary Cahill or Azpilicueta in the build-up to spread the play. 

\paragraph{Diego Costa:} Costa generates a small number of threats despite being a striker. This is expected as he is responsible for transforming the goalscoring chances rather than being at the origin of the danger. Moreover, his GoT$^{(ind)}_{90}$ statistic is particularly low since he has a low number of touches per time unit and many of his touches in the danger zone are not recorded in the constructed counting process.
 \\
 \noindent
 \\
We can clearly see that considering indirect contribution to threat generation is important for defenders and midfielders. These positions are generally at the base of the danger creation process. They have  small  GoT$^{(dir)}$ indices. However, indirect generated threat combined with the consideration of the number of touches allows us to effectively compare players playing in deeper positions.
 \\
 \noindent
 \\
From the graphical representation in Figure \ref{fig:graphchelsea}, we can identify some patterns that lead to a dangerous situation. When facing a team like Chelsea in the 2016-2017 season, some strategies can be derived from this analysis:
\begin{itemize}
\item As illustrated in Figure \ref{fig:graphchelsea}, the right side of Chelsea combines a lot for threat generation and should be disrupted at the root. Azpilicueta should be stopped from feeding the ball to the midfielders or directly to Pedro. 
\item The left side relies much more on the direct offensive output of Eden Hazard. In fact, all of Gary Cahill, Matic and Marcos Alonso mostly aim at delivering the ball to the left winger. To neutralize the threat of the left side, it is essential to prevent the ball from reaching Hazard. This can be achieved by marking him closely or by constantly closing the passing lanes to him.
\item  Goalkeeper Courtois is successful in targetting Marcos Alonso directly. This passing pattern should be considered when pressing Chelsea.
\end{itemize}

\section{Ligue 1 2021-22 season analysis}
\label{sec:ligue1}
In this section, we provide a ranking of players and teams from Ligue 1 in the 2021-2022 season based on their generation of threat. To maintain homogeneity, we only consider for each team the games where they use their main formation cluster, see Table \ref{tab:formationclusters} in Appendix \ref{app:ranking} for the list of formation clusters of each team.

\subsection{Generated threat to rank players in a position}

Each position on the pitch imposes a different role on the player who occupies it. In particular, we cannot expect the  same player to produce the same GoT metrics at two different positions. Therefore, we choose to evaluate players when they play in a particular position. This approach will also allow us to determine the optimal position for a player to maximize a GoT metric of interest. Additionally, we apply a filter to consider only players who play at least 600 minutes at a given position, with playing time calculated based on games in which the player features for at least 45 minutes.
 \\
 \noindent
 \\
Given a player and a position, we record the games in which the player occupies the position. The remaining positions may feature different players at each game. Whenever a player from his team is substituted, we do not consider the rest of the game in the construction of the counting process.  We fit a Hawkes process and assign to the player the generated threat indices of his position.  Tables \ref{tab:directdanger} and  \ref{tab:totaldanger} present the top twenty players in Ligue 1 in terms of GoT$^{(dir)}$ and GoT$^{(ind)}_{90}$,  respectively (see Tables \ref{tab:fullrankingdirect} and \ref{tab:fullrankingtotal} in Appendix \ref{app:ranking} for the Top 100). The tables also include Monte Carlo estimates of the standard error using $100$ samples. We display these two indices because they quantify the two extremes of the danger generation process. GoT$^{(dir)}$ isolates the direct impact of  players while GoT$^{(ind)}_{90}$ measures their participation in the chain of events leading to threats.
 \\
 \noindent
 \\
In reference to the results in Section \ref{subsec:toy}, one should keep in mind that the fewer minutes a player plays in a position, the less accurate the estimate of his generated threat is. Moreover, our estimation relies on selected games only. When a player has a limited number of minutes in a position, a good GoT metric should be interpreted as a measure of performance across the considered games only. For example, Moses Simon ranking third in GoT$^{(dir)}$ should not be surprising as he provided seven assists in the 1200 minutes but only gave one more assist in the remaining games when the team plays in a different formation or when he plays in a different position.
 \\
 \noindent
 \\
Below are some observations based on the results:

\begin{table}[t!]
    \centering
\resizebox{0.7\columnwidth}{!}{\begin{tabular}{llllll}
\toprule
\toprule
\bfseries Rank & \bfseries Name & \bfseries Position & \bfseries Team & \bfseries Minutes & \bfseries GoT$^{(dir)}$  {\tiny (SE) }\\
\midrule
1 & Lionel Messi & 10 & Paris Saint-Germain & 630 & 0.130  {\tiny (0.025)}\\
2 & Ángel Di María & 10 & Paris Saint-Germain & 1171 & 0.128  {\tiny (0.020)}\\
3 & Moses Simon & 11 & Nantes & 1222 & 0.120  {\tiny (0.024)}\\
\midrule
4 & Kylian Mbappé & 9 & Paris Saint-Germain & 1338 & 0.110  {\tiny (0.026)} \\
5 & Lionel Messi & 9 & Paris Saint-Germain & 675 & 0.109  {\tiny (0.023)} \\
6 & Martin Terrier & 11 & Rennes & 1386 & 0.108  {\tiny (0.017)} \\
\midrule
7 & Kylian Mbappé & 11 & Paris Saint-Germain & 1066 & 0.107  {\tiny (0.020)} \\
8 & Romain Faivre & 7 & Brest & 630 & 0.106  {\tiny (0.024)} \\
8 & Houssem Aouar & 7 & Lyon & 810 & 0.106  {\tiny (0.023)} \\
\midrule
10 & Sofiane Boufal & 9 & Angers & 771 & 0.100  {\tiny (0.020)} \\
11 & Jonathan Ikoné & 7 & Lille & 767 & 0.096  {\tiny (0.024)} \\
12 & Wissam Ben Yedder & 9 & Monaco & 1625 & 0.094  {\tiny (0.020)} \\
\midrule
13 & Franck Honorat & 11 & Brest & 838 & 0.093  {\tiny (0.021)} \\
13 & Karl Toko-Ekambi & 11 & Lyon & 1855 & 0.093  {\tiny (0.015)} \\
15 & Benjamin Bourigeaud & 10 & Rennes & 1719 & 0.092  {\tiny (0.014)}\\
\midrule
16 & Sofiane Boufal & 11 & Angers & 665 & 0.091  {\tiny (0.024)}\\
17 & Justin Kluivert & 11 & Nice & 1207 & 0.090  {\tiny (0.020)} \\
19 & Dimitri Payet & 9 & Marseille & 617 & 0.088  {\tiny (0.022)} \\
\midrule
19 & Kevin Gameiro & 10 & Strasbourg & 673 & 0.088  {\tiny (0.024)}\\
19 & Neymar & 11 & Paris Saint-Germain & 1258 & 0.088  {\tiny (0.014)} \\
\bottomrule
\bottomrule
\end{tabular}}
    
    \vspace{0.4em}
    \caption{Ranking of Ligue 1 players in terms of GoT$^{(dir)}$.}\label{tab:directdanger}
\end{table}

\paragraph{GoT$^{(dir)}$ vs GoT$^{(ind)}_{90}$:} GoT$^{(dir)}$ captures the intrinsic ability of a player to advance the ball to the opponent's danger area while GoT$^{(ind)}_{90}$ incorporates possible combinations with teammates. Therefore, the style of play and the ability of teammates can have an impact on the value of GoT$^{(ind)}_{90}$. These two indices describe different ways to contribute to threat generation and allow us to select different profiles of players. For example, the Paris Saint-Germain midfielder Verratti produces high values of GoT$^{(ind)}_{90}$ while Moses Simon from FC Nantes features in the top positions in terms of GoT$^{(dir)}$.

\paragraph{Jason Berthomier as a surprising pick:} In his only season in Ligue 1, Jason Berthomier delivered excellent values of GoT$^{(ind)}_{90}$. The Clermont Foot midfielder ranks 43$^{rd}$ in terms of GoT$^{(dir)}$ and climbs up to the tenth position in the  GoT$^{(ind)}_{90}$ ranking. This proves that he is consistently involved in the generation of dangerous situations for his team and is successful in feeding the forward players. 
\paragraph{Téji Savanier excels in midfield:} Téji Savanier stands out as an interior midfielder in the 433 formation of Montpellier. With eight goals and seven assists, it is no surprise that he is central to the process of threat generation of his team.  He ranks eighth in GoT$^{(ind)}_{90}$ and outperforms many offensive players in the league. This confirms the quality of Téji Savanier and his good performance during the 2021-2022 season.
\paragraph{A defender in the Top 20:} Frederic Guilbert of Strasbourg is a defender who excels at creating threats, ranking 18$^{th}$ in GoT$^{(ind)}_{90}$. In fact, his team deploys a 532 formation that provides enough cover for the fullbacks to play offensively.  The same holds for  Jonathan Clauss who acts almost as a right midfielder in the Lens formation and ranks 33$^{rd}$ in GoT$^{(ind)}_{90}$. This is also not surprising as Clauss ranks third in the league in the number of passes that lead to a shot, another proof of his creative play.
\paragraph{A good season from Messi in generated threat:} Despite underperforming in terms of scoring goals, Lionel Messi delivers outstanding values of generated threat both directly given his dribbling and passing quality, and indirectly given his involvement in ball possession. Additionally, we observe that his performance increases slightly when playing in his natural position as a 10 in the 433 formation. The right wing is Messi's best position as he poses more of a threat cutting inside from the right.
\paragraph{Optimal position for some players:} Romain Faivre stands out in both GoT$^{(dir)}$ and GoT$^{(ind)}_{90}$, ranking among the top twenty players. This is in fact expected because, when playing as a right midfielder in the 442 formation of Brest, the player performed well and was involved in six goals in just 660 minutes. Similarly, Houssem Aouar was successful as an interior midfielder in the 433 formation. He scored three and assisted three more in the considered period, earning him a top spot on our list.

\paragraph{A metric that does not value center forwards:} 
Very few strikers make the Top 20 in the two metrics. This is because the role of some center forwards is to receive the ball in the danger area and not necessarily to be at the origin of the threat. This is even more pronounced when looking at GoT$^{(ind)}_{90}$. For example, Mbappé, the top scorer in the league, barely makes it to the Top 20. Mbappe is not known for participating in possession and touching the ball a lot  but as an aggressive transition player. In contrast, midfielders such as Verratti and Guimarães, that are involved in the build-up of a lot of dangerous situations, feature in the top positions in terms of GoT$^{(ind)}_{90}$.

\begin{table}[t!]
    \centering
   \resizebox{0.7\columnwidth}{!}{
\begin{tabular}{llllll}
\toprule
\toprule
\bfseries Rank & \bfseries Name & \bfseries Position & \bfseries Team & \bfseries Minutes & \bfseries GoT$^{(ind)}_{90}$  {\tiny (SE) }\\
\midrule
1 & Lionel Messi & 10 & Paris Saint-Germain & 630 & 14.911   {\tiny (2.816)} \\
2 & Ángel Di María & 10 & Paris Saint-Germain & 1171 & 13.218   {\tiny (2.045)} \\
3 & Neymar & 11 & Paris Saint-Germain & 1258 & 12.724   {\tiny (1.709)} \\
\midrule
4 & Marco Verratti & 4 & Paris Saint-Germain & 602 & 12.581   {\tiny (2.655)} \\
5 & Lionel Messi & 9 & Paris Saint-Germain & 675 & 12.353   {\tiny (2.338)} \\
6 & Romain Faivre & 7 & Brest & 630 & 10.402   {\tiny (2.412)} \\
\midrule
7 & Houssem Aouar & 7 & Lyon & 810 & 10.077   {\tiny (2.014)} \\
8 & Téji Savanier & 7 & Montpellier & 2209 & 9.608   {\tiny (1.294)} \\
9 & Marco Verratti & 8 & Paris Saint-Germain & 1069 & 9.446   {\tiny (1.442)} \\
\midrule
10 & Jason Berthomier & 7 & Clermont & 1244 & 9.340   {\tiny (1.305)} \\
11 & Benjamin Bourigeaud & 10 & Rennes & 1719 & 9.211   {\tiny (1.228)} \\
12 & Sofiane Boufal & 9 & Angers & 771 & 9.100   {\tiny (1.799)} \\
\midrule
13 & Bruno Guimarães & 4 & Lyon & 900 & 8.817   {\tiny (1.995)} \\
14 & Dimitri Payet & 9 & Marseille & 617 & 8.815   {\tiny (2.021)} \\
15 & Moses Simon & 11 & Nantes & 1222 & 8.790   {\tiny (1.806)} \\
\midrule
16 & Martin Terrier & 11 & Rennes & 1386 & 8.639   {\tiny (1.335)} \\
17 & Kylian Mbappé & 11 & Paris Saint-Germain & 1066 & 8.577   {\tiny (1.657)} \\
18 & Frédéric Guilbert & 2 & Strasbourg & 2428 & 8.421   {\tiny (1.002)} \\
\midrule
19 & Ruben Aguilar & 2 & Monaco & 1205 & 8.019   {\tiny (1.417)} \\
20 & Lovro Majer & 7 & Rennes & 1302 & 7.927   {\tiny (1.321)} \\
\bottomrule
\bottomrule
\end{tabular}}
    \vspace{0.4em}
    \caption{Ranking of Ligue 1 players in terms of GoT$^{(ind)}_{90}$.}\label{tab:totaldanger}
\end{table}
\subsection{Ranking the central defenders' involvement in terms of GoT}
 To quantify the involvement of central defenders in danger creation, we use the indirect generation of threat per 90 minutes. This is because the direct generation of threat (GoT$^{(dir)}$) values are particularly low for defenders and therefore cannot be used to compare players. While GoT$^{(ind)}_{90}$ is influenced by the quality of the offensive players and team style of play, it also provides valuable information on the role of  defenders in the team's build-up scheme. For instance, a center-back who is technically proficient but avoids taking risks and does not contribute much to ball progression will have a low value of GoT$^{(ind)}_{90}$. This metric strikes a balance in measuring a player's intrinsic ability as well as their involvement within the team. Table \ref{tab:dangergk} displays the Top 10 best central defenders with the highest values of GoT$^{(ind)}_{90}$.
 \\
 \noindent
 \\
It is no surprise that Marquinhos and Kimpembe take the first two spots, given that they are part of Paris Saint-Germain, the most dominant team in Ligue 1. This is of course due to their technical ability, but there is also a factor due to the high possession values and danger creation ability of their team. The same holds for Nayef Aguerd and Warmed Omari that contribute significantly to ball progression, primarily through accurate long balls. The third-placed  is  Facundo Medina. The Lens defender is well known for his range of passing and for his ability to switch play from one side to the other. In particular, he ranks tenth in the league in terms of accurate passes per 90 minutes. William Saliba naturally completes the Top 5. The Marseille player excels with the ball at his feet and ranks third in accurate passing in Ligue 1. The player has now moved to Arsenal, a team that likes to play from the back, and continues to deliver in that aspect of the game. 

\begin{table}[t!]
    \centering
\resizebox{0.7\columnwidth}{!}{\begin{tabular}{llllll}
\toprule
\toprule
\bfseries Rank & \bfseries Name & \bfseries Position & \bfseries Team & \bfseries Minutes & \bfseries GoT$^{(ind)}_{90}$  {\tiny (SE) }\\
\midrule
1 & Marquinhos & 5 & Paris Saint-Germain & 2340 & 5.625  {\tiny (0.805) } \\
2 & Presnel Kimpembe & 6 & Paris Saint-Germain & 1840 & 5.230  {\tiny (0.747) }\\
3 & Facundo Medina & 4 & Lens & 1329 & 4.953 {\tiny (1.008) } \\
\midrule
4 & Nayef Aguerd & 6 & Rennes & 1698 & 4.908  {\tiny (0.694) }\\
5 & William Saliba & 5 & Marseille & 1800 & 4.652  {\tiny (0.756) }\\
6 & Jason Denayer & 6 & Lyon & 630 & 4.591 {\tiny (1.075) } \\
\midrule
7 & Jonathan Gradit & 6 & Lens & 1710 & 4.535  {\tiny (0.739) }\\
8 & Warmed Omari & 5 & Rennes & 1710 & 4.407  {\tiny (0.747) }\\
9 & Damien Da Silva & 5 & Lyon & 612 & 4.182 {\tiny (1.458) } \\
\midrule
10 & Dante & 6 & Nice & 2880 & 3.707 {\tiny (0.609) } \\
11 & Duje Caleta-Car & 6 & Marseille & 1397 & 3.462  {\tiny (0.715) }\\
12 & Lucas Perrin & 6 & Strasbourg & 2329 & 3.185  {\tiny (0.614) }\\
\midrule
13 & Kevin Danso & 5 & Lens & 1620 & 3.105  {\tiny (0.651) }\\
14 & Benoît Badiashile & 6 & Monaco & 975 & 3.075  {\tiny (0.762) }\\
15 & Castello Lukeba & 6 & Lyon & 1375 & 3.074  {\tiny (0.693) }\\
\midrule
16 & Mamadou Sakho & 6 & Montpellier & 1962 & 2.998  {\tiny (0.554) }\\
17 & Florent Ogier & 6 & Clermont & 2329 & 2.949  {\tiny (0.526) }\\
18 & Guillermo Maripán & 6 & Monaco & 810 & 2.916  {\tiny (0.989) }\\
\midrule
19 & Guillermo Maripán & 5 & Monaco & 605 & 2.915  {\tiny (1.486) }\\
20 & Jean-Clair Todibo & 5 & Nice & 3123 & 2.864  {\tiny (0.526) }\\
\bottomrule
\bottomrule
\end{tabular}}
    
    \vspace{0.4em}
    \caption{Ranking of Ligue 1 central defenders in terms of GoT$^{(ind)}_{90}$.}\label{tab:dangergk}
\end{table}

\subsection{GoT$^{(dir)}$ to rank teams}
To verify the consistency of our metrics, we rank Ligue 1 teams based on their aggregate values of GoT$^{(dir)}$. This metric can be considered as an indicator of squad quality. For each club, we fit a 12-dimensional Hawkes process to all matches in which they use their primary formation cluster, regardless of the players occupying each position. We then sum the estimated direct threat per touch GoT$^{(dir)}$ for all the positions.
 \\
 \noindent
 \\
Table \ref{tab:classementl1} shows the resulting Top 10 based on generated threat.  Our metric describes an important part of the offensive performance but obviously does not cover all aspects of the game. Nevertheless, it remains a very good measure of the quality of the team. Our ranking shows a significant $62\%$  Kendall correlation with the realized ranking of Ligue 1. This is achieved while only looking at ball touch and threat event timestamps to infer player abilities. Below are some observations from the ranking:

\begin{itemize}
    \item Rennes climbs to the second position in our ranking. This is because the team was very attack-minded in the 2021-2022 season and managed to score 82 goals, one of the highest totals in Europe. Their expected threat is proof of their offensive output.
    \item Olympique Lyonnais, ranked eighth in Ligue 1, still had a very prolific season offensively. They have the third-highest total of goals and the second-highest total of expected goals. It is therefore natural they are fourth with respect to our offensive metric. 
\end{itemize}

\begin{table}[t!]
    \centering
    \resizebox{0.5\columnwidth}{!}{\begin{tabular}{llll}
\toprule
\toprule
\bfseries Team & \bfseries GoT$^{(dir)}$ {\tiny (SE)}  & \bfseries Ligue 1 ranking & \bfseries Goals scored \\
\midrule
\bfseries Paris Saint-Germain & 0.42  {\tiny (0.024)}& 1 & 90 \\
\bfseries Rennes & 0.41  {\tiny (0.027)}& 4 & 82 \\
\bfseries Monaco & 0.41  {\tiny (0.035)}& 3 & 65 \\
\midrule
\bfseries Lyon & 0.36  {\tiny (0.025)}& 8 & 66 \\
\bfseries Marseille & 0.36  {\tiny (0.033)}& 2 & 63 \\
\bfseries Lens & 0.30  {\tiny (0.029)}& 7 & 62 \\
\midrule
\bfseries Nice & 0.28  {\tiny (0.023)}& 5 & 52 \\
\bfseries Strasbourg & 0.26  {\tiny (0.021)}& 6 & 60 \\
\bfseries Lille & 0.26  {\tiny (0.022)}& 10 & 48 \\
\midrule
\bfseries Reims & 0.26  {\tiny (0.035)}& 12 & 43 \\
\bottomrule
\bottomrule
\end{tabular}}
\vspace{0.3em}
    \caption{Top 10 Ligue 1 teams with respect to aggregated GoT$^{(dir)}$ of starting eleven.}
    \label{tab:classementl1}
\end{table}

\section{Conclusion and future work}
In order to measure a player's ability to create threat in football, we develop model-based metrics that rely on Hawkes processes. These processes provide an easy to interpret way to capture hierarchy between event times. Thanks to this modeling, we are able to identify the players whose touches are most consistently correlated with subsequent threats. We derive four different metrics each describing different ways to create danger. On the one hand, the direct generation of threat metrics GoT$^{(dir)}$ and GoT$^{(dir)}_{90}$ allow us to isolate the intrinsic ability of players. On the other hand, GoT$^{(ind)}$ 
 and GoT$^{(ind)}_{90}$ indicate the indirect contribution to the generation of threat through interactions with other positions. Beyond crediting players for danger generation, our approach can also be used to  quantify and visualize the synergies between players on the pitch and identify the patterns that lead to dangerous situations.
 \\
 \noindent
 \\
We demonstrate our methodology can successfully detect and rank the key players in the 2021-2022 Ligue 1 season, who contribute to their team's offensive output. The results we find are consistent with the observed performances of the retrieved players, but also reveal some surprising choices. Through the example of Chelsea in the 2016-2017 season, we show that our model-based approach can help teams make data-driven decisions about their tactics. By primarily looking at timestamps of ball touches, we gain a deeper understanding of the threat generation process of a team.
 \\
 \noindent
 \\
 Our methodology can be adapted to study the dynamics of other sports that have available event data. This is particularly valid for sports where the set of players on the pitch is relatively stable like football. This stability allows us to construct counting processes where each component consistently corresponds to the same player. Applying this approach to basketball, for example, presents additional challenges due to the high frequency of substitutions. To address this, one strategy is to design a point process where each component tracks the events of a specific role (point guard, shooting guard, small forward, power forward, center) regardless of the player's identity. This approach is similar to the analysis presented in Section~\ref{app:rennes} of the Appendix. Metrics estimated for each role can then be attributed to the player occupying that position the most during the game. Furthermore, a notable advantage in basketball is that the frequent scoring events can be used to represent threats.
 \\
 \noindent
 \\
Future work will include  exploring the application of our model-based metrics for optimal team selection. In fact, if we are capable of inferring the branching matrix parameters linking players from different teams, we can measure the impact of a potential transfer on the danger creation process. In addition, we can use this framework to capture interactions of players with other game states different from threats. In particular, by replacing the threat events with ball losses, we can effectively analyze the defensive aspect of the game and determine players whose touches are most correlated with a turnover. 

\paragraph{Acknowledgment:} The authors thank Anna Bonnet for her help with the estimation of Hawkes processes in large dimensions. They are also grateful to Charlotte Dion and Céline Duval. The authors gratefully acknowledge financial support from the chairs  “Machine Learning \& Systematic Methods in Finance” and “Deep Finance and Statistics”.

\newpage
\bibliographystyle{apalike}
\bibliography{references}

\begin{thebibliography}{}

\bibitem[Adamopoulos, 1976]{adamopoulos1976cluster}
Adamopoulos, L. (1976).
\newblock Cluster models for earthquakes: Regional comparisons.
\newblock {\em Journal of the International Association for Mathematical Geology}, 8:463--475.

\bibitem[Bonnet et~al., 2022a]{bonnet2022neuronal}
Bonnet, A., Dion-Blanc, C., Gindraud, F., and Lemler, S. (2022a).
\newblock {Neuronal network inference and membrane potential model using multivariate Hawkes processes}.
\newblock {\em Journal of Neuroscience Methods}, 372:109550.

\bibitem[Bonnet et~al., 2022b]{bonnet2022inference}
Bonnet, A., Herrera, M.~M., and Sangnier, M. (2022b).
\newblock {Inference of multivariate exponential Hawkes processes with inhibition and application to neuronal activity}.
\newblock {\em arXiv preprint arXiv:2205.04107}.

\bibitem[Br{\'e}maud, 1981]{bremaud1981point}
Br{\'e}maud, P. (1981).
\newblock {\em Point processes and queues: martingale dynamics}, volume~50.
\newblock Springer.

\bibitem[Green, 2012]{xgblog}
Green, S. (2012).
\newblock {Assessing the performance of Premier League goalscorers}.
\newblock \url{https://www.statsperform.com/resource/assessing-the-performance-of-premier-league-goalscorers/}.

\bibitem[Hawkes, 1971a]{hawkes1971point}
Hawkes, A.~G. (1971a).
\newblock Point spectra of some mutually exciting point processes.
\newblock {\em Journal of the Royal Statistical Society: Series B (Methodological)}, 33(3):438--443.

\bibitem[Hawkes, 1971b]{hawkes1971spectra}
Hawkes, A.~G. (1971b).
\newblock Spectra of some self-exciting and mutually exciting point processes.
\newblock {\em Biometrika}, 58(1):83--90.

\bibitem[Hawkes and Oakes, 1974]{hawkes1974cluster}
Hawkes, A.~G. and Oakes, D. (1974).
\newblock A cluster process representation of a self-exciting process.
\newblock {\em Journal of Applied Probability}, 11(3):493--503.

\bibitem[Jaisson and Rosenbaum, 2015]{jaisson2015limit}
Jaisson, T. and Rosenbaum, M. (2015).
\newblock {Limit theorems for nearly unstable Hawkes processes}.
\newblock {\em The Annals of Applied Probability}, 25(2).

\bibitem[Lambert et~al., 2018]{lambert2018reconstructing}
Lambert, R.~C., Tuleau-Malot, C., Bessaih, T., Rivoirard, V., Bouret, Y., Leresche, N., and Reynaud-Bouret, P. (2018).
\newblock {Reconstructing the functional connectivity of multiple spike trains using Hawkes models}.
\newblock {\em Journal of neuroscience methods}, 297:9--21.

\bibitem[Ogata, 1981]{ogata1981lewis}
Ogata, Y. (1981).
\newblock On lewis' simulation method for point processes.
\newblock {\em IEEE transactions on information theory}, 27(1):23--31.

\bibitem[Ogata, 1988]{ogata1988statistical}
Ogata, Y. (1988).
\newblock {Statistical models for earthquake occurrences and residual analysis for point processes}.
\newblock {\em Journal of the American Statistical association}, 83(401):9--27.

\bibitem[Ogata et~al., 1978]{ogata1978estimators}
Ogata, Y. et~al. (1978).
\newblock The asymptotic behaviour of maximum likelihood estimators for stationary point processes.
\newblock {\em Annals of the Institute of Statistical Mathematics}, 30(1):243--261.

\bibitem[Singh, 2018]{xtblog}
Singh, K. (2018).
\newblock {Introducing Expected Threat (xT)}.
\newblock \url{https://karun.in/blog/expected-threat.html}.

\bibitem[Whitmore, 2021]{xablog}
Whitmore, J. (2021).
\newblock {What Are Expected Assists (xA)?}
\newblock \url{https://theanalyst.com/eu/2021/03/what-are-expected-assists-xa/}.

\end{thebibliography}

\newpage

\begin{center}
{\Large \bf {\centering Appendix}}
\end{center}

\appendix
\label{appendix}
\section{Stade Rennais}
\label{app:rennes}
In this appendix, we present a detailed analysis of one of the Ligue 1 teams in the 2021-2022 season.

\subsection{Selected games}

In the same spirit as Section \ref{sec:chelsea}, we choose a collection of games where the formation is the same and starting lineup is as stable as possible. Table \ref{tab:selectedrennes} shows the selected matches for Stade Rennais. The team plays in a 433 formation in all of these games but the starting eleven is not always exactly the same. In fact, some players are sometimes rotated for a game or two,  but we assume that the substitute behaves approximately the same as the starting player. Stade Rennais line up as follows in the selected games, where the main player in each position is in bold:
\begin{itemize}
    \item \textbf{Gomis}/Alemdar
\item \textbf{Traore} - \textbf{Omari}/Bade - \textbf{Aguerd}/Bade/Santamaria - \textbf{Truffert}/Meling
\item \textbf{Majer} - \textbf{Santamaria}/Martin - \textbf{Tait}
\item \textbf{Bourigeaud} - \textbf{Laborde}/Guirassy - \textbf{Terrier}
\end{itemize}
\noindent
We construct a 12-dimensional counting process from the selected Stade Rennais games regardless of the players starting. We use the data from each game as long as the eleven players on the pitch correspond to the scheme provided above. We then fit a 12-dimensional Hawkes process and associate the estimated metrics of each position with the main player occupying it.

\begin{table}[t]
    \centering
     \resizebox{0.6\columnwidth}{!}{\begin{tabular}{llll}
    \toprule
    \toprule
        \textbf{Date} & \textbf{Opponent} & \textbf{Home or Away} & \textbf{Competition} \\ 
        \midrule
        May 11, 2022 & Nantes & Away & French Ligue 1 \\ 
        Apr 2, 2022 & Nice & Away & French Ligue 1 \\ 
        May 14, 2022 & Marseille & Home & French Ligue 1 \\ 
        \midrule
        Dec 22, 2021 & Monaco & Away & French Ligue 1 \\ 
        Mar 20, 2022 & Metz & Home & French Ligue 1 \\ 
        Apr 15, 2022 & Monaco & Home & French Ligue 1 \\ 
        \midrule
        Apr 30, 2022 & St Etienne & Home & French Ligue 1 \\ 
        Apr 24, 2022 & Lorient & Home & French Ligue 1 \\ 
        Nov 20, 2021 & Montpellier & Home & French Ligue 1 \\ 
        \midrule
        May 21, 2022 & Lille & Away & French Ligue 1 \\ 
        Nov 7, 2021 & Lyon & Home & French Ligue 1 \\ 
        \bottomrule
        \bottomrule
    \end{tabular}}
    \vspace{0.5em}
    \caption{List of selected games for Stade Rennais F.C.}
    \label{tab:selectedrennes}
\end{table}

\subsection{Results and discussion}
\begin{figure}[t]
    \centering
    \begin{minipage}{0.9\textwidth}
        \centering
        \scalebox{0.6}{\includegraphics{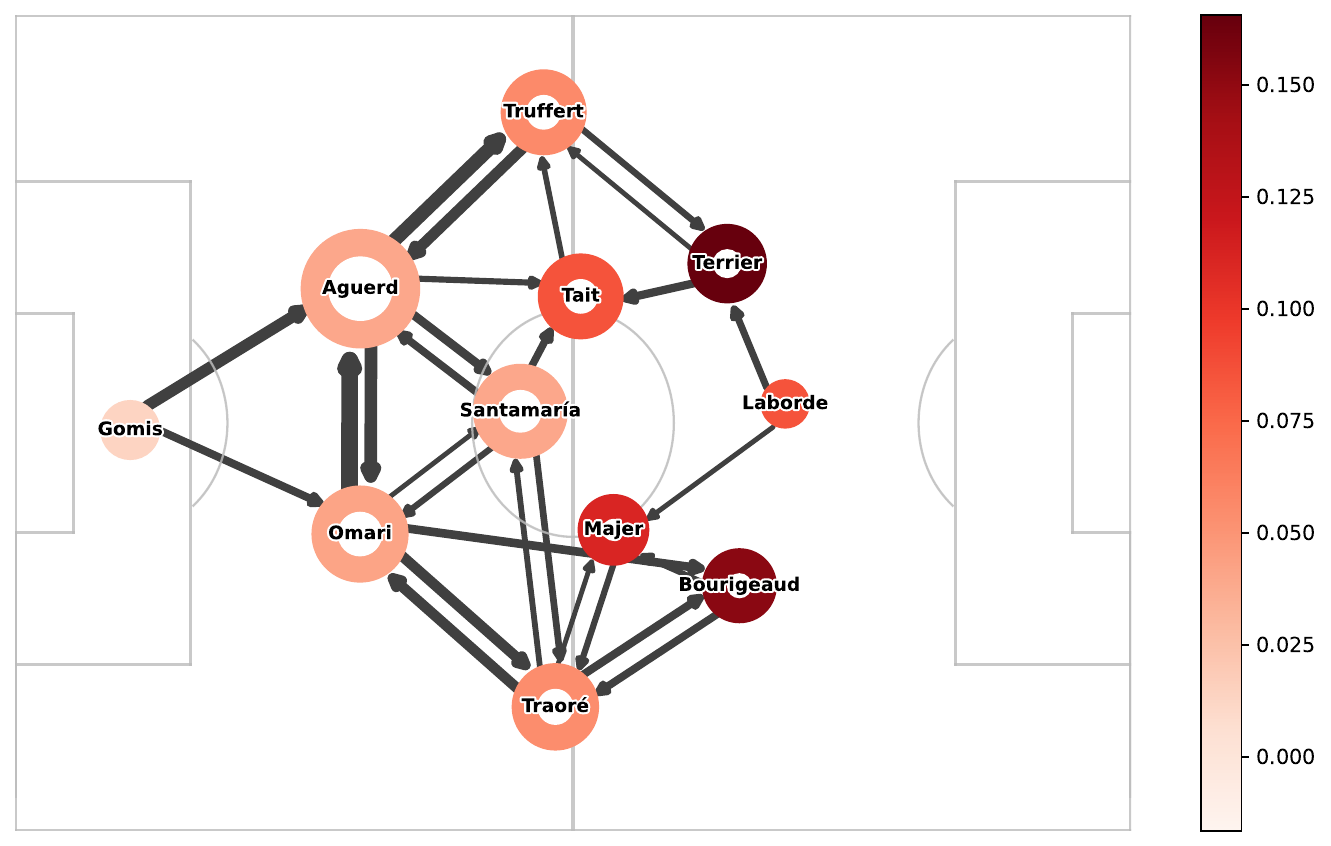}}\vspace*{-.5em}
        \captionof{figure}{Graph summarizing the interactions between Stade Rennais players. The width of an arrow from player $p_1$ to player $p_2$ is proportional to the expected number of touches of player $p_2$ generated by one touch from player $p_1$. The size of the circle of player $p$ is proportional to the sum of the arrow sizes received, indicating the involvement of the player in the considered games. The color of the circle represents the GoT$^{i}$ index for each player.}\label{fig:graphrennes}      
    \end{minipage}
\end{figure}

\begin{figure}[t]
    \centering
    \begin{minipage}{0.9\textwidth}
        \centering
        \scalebox{0.6}{\includegraphics{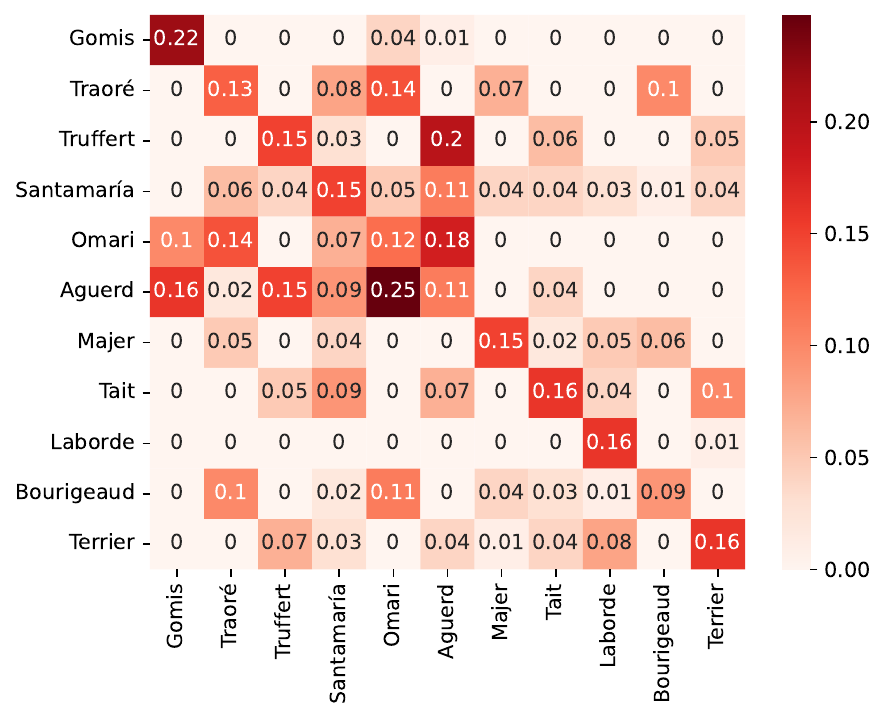}}\vspace*{-.5em}
        \captionof{figure}{Estimated branching matrix for Stade Rennais.}\label{fig:bmrennes}      
    \end{minipage}
    
\end{figure}

In Table \ref{tab:dangerrennes}, we rank the Stade Rennais players with respect to generated threat metrics. Figure \ref{fig:graphrennes} graphically represents  the direct interactions between them and  Figure \ref{fig:bmrennes} displays the estimated branching matrix. 
\noindent
We can see that the team adopts a 433 shape that progresses mainly through the wings. The danger creation is asymmetric with more combinations occurring on the right side, where Majer is the most creative midfielder. Interestingly, despite being a central midfielder, Flavien Tait delivers a large value of GoT$^{(ind)}_{90}$, indicating that he is a significant contributor to the team's offensive efforts. In contrast, although Santamaria has more possession, he has limited involvement in creating threats. This difference in their threat generation can be attributed to their distinct roles on the field. On one hand, Tait is a more box-to-box midfielder who frequently projects forward and has a considerable direct threat metric. On the other hand, Santamaria belongs to a class of defensive midfielders who act as anchor points. They participate in the buildup close to the center backs and have limited interactions with the forward positions.
 \\
 \noindent
 \\
The main threat sources are Bourigeaud, Majer, and Terrier. These three players are outstanding going forward. Terrier is the leader of the team in goalscoring and ranks third in Ligue 1 but seems to be involved in danger creation as well. Bourigeaud generating the most threat is not surprising since he is the creative force of the team. In fact, he ranks first in the league in terms of key passes with 3.2 per game, and first in accurate crosses with 104 in the season.  
 \\
 \noindent
 \\
As expected, the center backs have zero direct threat contribution. However, in terms of indirect threat per 90 minutes GoT$^{(ind)}_{90}$, Aguerd and Omari rank fourth and sixth in the team respectively. The pair generates danger through their involvement  in team build-up and possession. In particular, Aguerd and Omari are comfortable with the ball at their feet and rank eighth and twentieth in the league, respectively, in the number of passes per game with high success rates.
\begin{table}[!ht]
    \centering
    \resizebox{0.6\columnwidth}{!}{
\begin{tabular}{lllll}
\toprule
\toprule
\bfseries Player name & \bfseries GoT$^{(dir)}$ & \bfseries GoT$^{(ind)}$ & \bfseries GoT$^{(dir)}_{90}$ & \bfseries GoT$^{(ind)}_{90}$ \\
\midrule
\bfseries Benjamin Bourigeaud & 0.14 & 0.16 & 11.8 & 12.6 \\
\bfseries Martin  Terrier & 0.13 & 0.17 & 8.6 & 9.5 \\
\bfseries Lovro Majer & 0.08 & 0.11 & 6.9 & 8.1 \\
\midrule
\bfseries Flavien Tait & 0.06 & 0.09 & 5.5 & 7.2 \\
\bfseries Adrien Truffert & 0.02 & 0.06 & 2.4 & 5.2 \\
\bfseries Hamari Traoré & 0.01 & 0.05 & 1.5 & 4.6 \\
\midrule
\bfseries Nayef Aguerd & 0.00 & 0.04 & 0.0 & 4.5 \\
\bfseries Baptiste Santamaría & 0.00 & 0.05 & 0.4 & 4.3 \\
\bfseries Warmed Omari & 0.00 & 0.04 & 0.0 & 4.0 \\
\midrule
\bfseries Gaëtan Laborde & 0.04 & 0.08 & 2.0 & 3.0 \\
\bfseries Alfred Gomis & 0.00 & 0.01 & 0.0 & 0.7 \\
\bottomrule
\bottomrule
\end{tabular}}
\vspace{0.3em}
        \caption{Generated threat metrics for the players of Stade Rennais. The table is sorted by GoT$^{(ind)}_{90}$.}
    \label{tab:dangerrennes}
\end{table}

\noindent
Finally, we can observe from Figure \ref{fig:graphrennes} some remarkable circuits that lead to dangerous situations. These patterns of play should be taken into account by an opposing team when facing Stade Rennais: 
\begin{itemize}
    \item Aguerd $\rightarrow$ Truffert $\rightarrow$ Terrier $\rightarrow$ Threat.
    \item Terrier $\rightarrow$ Tait $\rightarrow$ Threat. Terrier is highly effective in generating direct threats, but he also frequently combines with Flavien Tait to create danger. Similarly, Bourigeaud often gives the ball to Lovro Majer to generate indirect threat.
    \item Omari $\rightarrow$ Traoré $\rightarrow$ Bourigeaud $\rightarrow$ Threat.
    \item Omari $\rightarrow$ Bourigeaud $\rightarrow$ Threat. This is a straightforward pattern from defense to attack that should be controlled. Omari is highly successful in progressing the ball, both through slow build-up play by passing the ball to the right-back Traoré, as well as through fast transitions with direct passes to Bourigeaud.
\end{itemize}

\section{Top 100 ranking of Ligue 1 player in terms of GoT}
\label{app:ranking}
\begin{table}[!ht]
    \centering
    \begin{minipage}{0.47\textwidth}
    \resizebox{\columnwidth}{0.3\textheight}{
\begin{tabular}{llllll}
\toprule
\toprule
\bfseries Rank & \bfseries Name & \bfseries Position & \bfseries Team & \bfseries Minutes & \bfseries GoT$^{(dir)}$\\
\midrule
1 & Lionel Messi & 10 & Paris Saint-Germain & 630 & 0.130 \\
2 & Ángel Di María & 10 & Paris Saint-Germain & 1171 & 0.128 \\
3 & Moses Simon & 11 & Nantes & 1222 & 0.120 \\
\midrule
4 & Kylian Mbappé & 9 & Paris Saint-Germain & 1338 & 0.110 \\
5 & Lionel Messi & 9 & Paris Saint-Germain & 675 & 0.109 \\
6 & Martin Terrier & 11 & Rennes & 1386 & 0.108 \\
\midrule
7 & Kylian Mbappé & 11 & Paris Saint-Germain & 1066 & 0.107 \\
8 & Romain Faivre & 7 & Brest & 630 & 0.106 \\
8 & Houssem Aouar & 7 & Lyon & 810 & 0.106 \\
\midrule
10 & Sofiane Boufal & 9 & Angers & 771 & 0.100 \\
11 & Jonathan Ikoné & 7 & Lille & 767 & 0.096 \\
12 & Wissam Ben Yedder & 9 & Monaco & 1625 & 0.094 \\
\midrule
13 & Franck Honorat & 11 & Brest & 838 & 0.093 \\
13 & Karl Toko-Ekambi & 11 & Lyon & 1855 & 0.093 \\
15 & Benjamin Bourigeaud & 10 & Rennes & 1719 & 0.092 \\
\midrule
16 & Sofiane Boufal & 11 & Angers & 665 & 0.091 \\
17 & Justin Kluivert & 11 & Nice & 1207 & 0.090 \\
19 & Dimitri Payet & 9 & Marseille & 617 & 0.088 \\
\midrule
19 & Kevin Gameiro & 10 & Strasbourg & 673 & 0.088 \\
19 & Neymar & 11 & Paris Saint-Germain & 1258 & 0.088 \\
21 & Jodel Dossou & 10 & Clermont & 1762 & 0.086 \\
\midrule
22 & Lucas Da Cunha & 10 & Clermont & 606 & 0.084 \\
22 & Jim Allevinah & 11 & Clermont & 858 & 0.084 \\
24 & Frédéric Guilbert & 2 & Strasbourg & 2428 & 0.082 \\
\midrule
25 & Armand Laurienté & 9 & Lorient & 842 & 0.080 \\
26 & Ludovic Blas & 7 & Nantes & 810 & 0.078 \\
27 & Gaël Kakuta & 9 & Lens & 976 & 0.077 \\
\midrule
28 & Arnaud Kalimuendo-Muinga & 11 & Lens & 724 & 0.075 \\
29 & Cengiz Ünder & 10 & Marseille & 1047 & 0.074 \\
29 & Florent Mollet & 10 & Montpellier & 1269 & 0.074 \\
\midrule
31 & Téji Savanier & 7 & Montpellier & 2209 & 0.071 \\
32 & Ghislain Konan & 3 & Reims & 1007 & 0.070 \\
33 & Lovro Majer & 7 & Rennes & 1302 & 0.068 \\
\midrule
34 & Kevin Volland & 7 & Monaco & 1131 & 0.067 \\
34 & Jonathan Clauss & 2 & Lens & 1940 & 0.067 \\
36 & Angelo Fulgini & 8 & Angers & 630 & 0.066 \\
\midrule
36 & Andy Delort & 10 & Nice & 1478 & 0.066 \\
38 & Javairô Dilrosun & 9 & Bordeaux & 675 & 0.064 \\
39 & Vanderson & 10 & Monaco & 619 & 0.061 \\
\midrule
40 & Lucas Paquetá & 7 & Lyon & 1248 & 0.060 \\
40 & Burak Yilmaz & 9 & Lille & 1900 & 0.060 \\
43 & Jonathan Bamba & 11 & Lille & 1763 & 0.059 \\
\midrule
43 & Thomas Foket & 2 & Reims & 631 & 0.059 \\
43 & Jason Berthomier & 7 & Clermont & 1244 & 0.059 \\
45 & Amine Gouiri & 9 & Nice & 1749 & 0.058 \\
\midrule
46 & Gaëtan Laborde & 9 & Rennes & 1305 & 0.057 \\
47 & Ibrahima Sissoko & 7 & Strasbourg & 826 & 0.055 \\
48 & Jonathan David & 10 & Lille & 2072 & 0.054 \\
\midrule
49 & Dimitri Lienard & 3 & Strasbourg & 1728 & 0.050 \\
51 & Flavien Tait & 8 & Rennes & 1129 & 0.046 \\
\bottomrule
\bottomrule
\end{tabular}
    
    }
    \end{minipage}
    \begin{minipage}{0.47\textwidth}

\resizebox{\columnwidth}{0.3\textheight}{
\begin{tabular}{llllll}
\toprule
\toprule
\bfseries Rank & \bfseries Name & \bfseries Position & \bfseries Team & \bfseries Minutes & \bfseries GoT$^{(dir)}$\\
\midrule
51 & Florian Sotoca & 10 & Lens & 1119 & 0.046 \\
51 & Jérémy Le Douaron & 10 & Brest & 759 & 0.046 \\
53 & Andy Delort & 9 & Nice & 795 & 0.045 \\
\midrule
55 & Youcef Atal & 2 & Nice & 1032 & 0.044 \\
55 & Randal Kolo Muani & 9 & Nantes & 876 & 0.044 \\
55 & Stephy Mavididi & 11 & Montpellier & 1585 & 0.044 \\
\midrule
55 & Aleksandr Golovin & 11 & Monaco & 607 & 0.044 \\
58 & Elbasan Rashani & 11 & Clermont & 1588 & 0.043 \\
59 & Mohamed Bayo & 9 & Clermont & 2331 & 0.042 \\
\midrule
60 & Abdu Conté & 3 & Troyes & 695 & 0.041 \\
60 & Sanjin Prcic & 11 & Strasbourg & 652 & 0.041 \\
64 & Bruno Guimarães & 4 & Lyon & 900 & 0.040 \\
\midrule
64 & Anthony Caci & 3 & Strasbourg & 1400 & 0.040 \\
64 & Kevin Gameiro & 9 & Strasbourg & 1458 & 0.040 \\
64 & Hicham Boudaoui & 7 & Nice & 1304 & 0.040 \\
\midrule
64 & Gerson & 8 & Marseille & 704 & 0.040 \\
67 & Sofiane Diop & 11 & Monaco & 631 & 0.039 \\
68 & Igor Silva & 2 & Lorient & 1197 & 0.038 \\
\midrule
69 & Renato Sanches & 8 & Lille & 951 & 0.037 \\
69 & Issa Kaboré & 2 & Troyes & 1679 & 0.037 \\
71 & Mohamed-Ali Cho & 10 & Angers & 807 & 0.035 \\
\midrule
71 & Angelo Fulgini & 9 & Angers & 751 & 0.035 \\
74 & Akim Zedadka & 2 & Clermont & 3330 & 0.034 \\
74 & Pol Lirola & 2 & Marseille & 863 & 0.034 \\
\midrule
74 & Adrien Thomasson & 7 & Strasbourg & 1853 & 0.034 \\
76 & Habib Diallo & 10 & Strasbourg & 859 & 0.033 \\
76 & Xavier Chavalerin & 11 & Troyes & 949 & 0.033 \\
\midrule
78 & Jean-Ricner Bellegarde & 11 & Strasbourg & 1454 & 0.032 \\
78 & Maxence Caqueret & 8 & Lyon & 1389 & 0.032 \\
80 & Vital N'Simba & 3 & Clermont & 2731 & 0.031 \\
\midrule
80 & Ruben Aguilar & 2 & Monaco & 1205 & 0.031 \\
83 & Seko Fofana & 8 & Lens & 1861 & 0.030 \\
83 & Ismail Jakobs & 3 & Monaco & 650 & 0.030 \\
\midrule
83 & Terem Moffi & 10 & Lorient & 911 & 0.030 \\
87 & Valère Germain & 9 & Montpellier & 1083 & 0.029 \\
87 & Stéphane Bahoken & 10 & Angers & 657 & 0.029 \\
\midrule
87 & Youssouf Fofana & 8 & Monaco & 1116 & 0.029 \\
87 & Mattéo Guendouzi & 7 & Marseille & 1350 & 0.029 \\
87 & Ludovic Ajorque & 10 & Strasbourg & 1334 & 0.029 \\
\midrule
90 & Ludovic Ajorque & 9 & Strasbourg & 1243 & 0.028 \\
90 & Marco Verratti & 4 & Paris Saint-Germain & 602 & 0.028 \\
93 & Baptiste Santamaría & 8 & Rennes & 675 & 0.027 \\
\midrule
93 & Vincent Le Goff & 3 & Lorient & 1440 & 0.027 \\
93 & Ricardo Mangas & 3 & Bordeaux & 613 & 0.027 \\
96 & Caio Henrique & 3 & Monaco & 1454 & 0.026 \\
\midrule
96 & Mihailo Ristic & 3 & Montpellier & 1150 & 0.026 \\
96 & Junior Sambia & 2 & Montpellier & 732 & 0.026 \\
98 & Souleyman Doumbia & 3 & Angers & 1797 & 0.025 \\
\midrule
98 & Przemyslaw Frankowski & 3 & Lens & 1191 & 0.025 \\
100 & Florian Tardieu & 8 & Troyes & 1530 & 0.024 \\
\bottomrule
\bottomrule
\end{tabular}

}
    \end{minipage}
    \vspace{0.4em}
    \caption{Ranking of Ligue 1 players in terms of GoT$^{(dir)}$.}\label{tab:fullrankingdirect}
\end{table}

\newpage
\begin{table}[!ht]
    \centering
    \begin{minipage}{0.47\textwidth}
    \resizebox{\columnwidth}{0.3\textheight}{
\begin{tabular}{llllll}
\toprule
\toprule
\bfseries Rank & \bfseries Name & \bfseries Position & \bfseries Team & \bfseries Minutes & \bfseries GoT$^{(ind)}_{90}$  \\
\midrule
1 & Lionel Messi & 10 & Paris Saint-Germain & 630 & 14.911 \\
2 & Ángel Di María & 10 & Paris Saint-Germain & 1171 & 13.218 \\
3 & Neymar & 11 & Paris Saint-Germain & 1258 & 12.724 \\
\midrule
4 & Marco Verratti & 4 & Paris Saint-Germain & 602 & 12.581 \\
5 & Lionel Messi & 9 & Paris Saint-Germain & 675 & 12.353 \\
6 & Romain Faivre & 7 & Brest & 630 & 10.402 \\
\midrule
7 & Houssem Aouar & 7 & Lyon & 810 & 10.077 \\
8 & Téji Savanier & 7 & Montpellier & 2209 & 9.608 \\
9 & Marco Verratti & 8 & Paris Saint-Germain & 1069 & 9.446 \\
\midrule
10 & Jason Berthomier & 7 & Clermont & 1244 & 9.340 \\
11 & Benjamin Bourigeaud & 10 & Rennes & 1719 & 9.211 \\
12 & Sofiane Boufal & 9 & Angers & 771 & 9.100 \\
\midrule
13 & Bruno Guimarães & 4 & Lyon & 900 & 8.817 \\
14 & Dimitri Payet & 9 & Marseille & 617 & 8.815 \\
15 & Moses Simon & 11 & Nantes & 1222 & 8.790 \\
\midrule
16 & Martin Terrier & 11 & Rennes & 1386 & 8.639 \\
17 & Kylian Mbappé & 11 & Paris Saint-Germain & 1066 & 8.577 \\
18 & Frédéric Guilbert & 2 & Strasbourg & 2428 & 8.421 \\
\midrule
19 & Ruben Aguilar & 2 & Monaco & 1205 & 8.019 \\
20 & Lovro Majer & 7 & Rennes & 1302 & 7.927 \\
21 & Lucas Da Cunha & 10 & Clermont & 606 & 7.882 \\
\midrule
22 & Kylian Mbappé & 9 & Paris Saint-Germain & 1338 & 7.733 \\
23 & Ghislain Konan & 3 & Reims & 1007 & 7.721 \\
24 & Sanjin Prcic & 11 & Strasbourg & 652 & 7.624 \\
\midrule
25 & Sofiane Boufal & 11 & Angers & 665 & 7.547 \\
26 & Lucas Paquetá & 7 & Lyon & 1248 & 7.450 \\
27 & Karl Toko-Ekambi & 11 & Lyon & 1855 & 7.365 \\
\midrule
28 & Jonathan Ikoné & 7 & Lille & 767 & 7.285 \\
29 & Franck Honorat & 11 & Brest & 838 & 7.207 \\
30 & Achraf Hakimi & 2 & Paris Saint-Germain & 1781 & 7.125 \\
\midrule
31 & Idrissa Gueye & 8 & Paris Saint-Germain & 662 & 7.090 \\
32 & Dimitri Lienard & 3 & Strasbourg & 1728 & 7.050 \\
33 & Gerson & 8 & Marseille & 704 & 7.016 \\
\midrule
34 & Vanderson & 10 & Monaco & 619 & 6.939 \\
35 & Ibrahima Sissoko & 7 & Strasbourg & 826 & 6.892 \\
36 & Ludovic Blas & 7 & Nantes & 810 & 6.816 \\
\midrule
37 & Gaël Kakuta & 9 & Lens & 976 & 6.802 \\
38 & Jonathan Clauss & 2 & Lens & 1940 & 6.757 \\
39 & Justin Kluivert & 11 & Nice & 1207 & 6.700 \\
\midrule
40 & Flavien Tait & 8 & Rennes & 1129 & 6.671 \\
41 & Kevin Gameiro & 10 & Strasbourg & 673 & 6.569 \\
42 & Florent Mollet & 10 & Montpellier & 1269 & 6.465 \\
\midrule
43 & Angelo Fulgini & 8 & Angers & 630 & 6.459 \\
44 & Renato Sanches & 8 & Lille & 951 & 6.427 \\
45 & Henrique & 3 & Lyon & 619 & 6.385 \\
\midrule
46 & Danilo Pereira & 4 & Paris Saint-Germain & 879 & 6.346 \\
47 & Pol Lirola & 2 & Marseille & 863 & 6.327 \\
48 & Jim Allevinah & 11 & Clermont & 858 & 6.303 \\
\midrule
49 & Youcef Atal & 2 & Nice & 1032 & 6.258 \\
50 & Maxence Caqueret & 8 & Lyon & 1389 & 6.229 \\
\bottomrule
\bottomrule
\end{tabular}
    
    }
    \end{minipage}
    \begin{minipage}{0.47\textwidth}

\resizebox{\columnwidth}{0.3\textheight}{
\begin{tabular}{llllll}
\toprule
\toprule
\bfseries Rank & \bfseries Name & \bfseries Position & \bfseries Team & \bfseries Minutes & \bfseries GoT$^{(ind)}_{90}$   \\
\midrule
51 & Vital N'Simba & 3 & Clermont & 2731 & 6.066 \\
52 & Anthony Caci & 3 & Strasbourg & 1400 & 6.023 \\
53 & Emerson & 3 & Lyon & 1848 & 6.019 \\
\midrule
54 & Aleksandr Golovin & 11 & Monaco & 607 & 5.973 \\
55 & Birger Meling & 3 & Rennes & 776 & 5.936 \\
56 & Juan Bernat & 3 & Paris Saint-Germain & 777 & 5.929 \\
\midrule
57 & Jodel Dossou & 10 & Clermont & 1762 & 5.864 \\
58 & Caio Henrique & 3 & Monaco & 1454 & 5.855 \\
59 & Jonas Martin & 4 & Rennes & 1115 & 5.810 \\
\midrule
60 & Thomas Foket & 2 & Reims & 631 & 5.805 \\
61 & Jonathan Bamba & 11 & Lille & 1763 & 5.632 \\
62 & Marquinhos & 5 & Paris Saint-Germain & 2340 & 5.625 \\
\midrule
63 & Florian Sotoca & 10 & Lens & 1119 & 5.598 \\
64 & Jordan Ferri & 8 & Montpellier & 2129 & 5.516 \\
65 & Aurélien Tchouaméni & 4 & Monaco & 1620 & 5.431 \\
\midrule
66 & Malo Gusto & 2 & Lyon & 1369 & 5.382 \\
67 & Cheick Oumar Doucouré & 7 & Lens & 1350 & 5.372 \\
68 & Armand Laurienté & 9 & Lorient & 842 & 5.345 \\
\midrule
69 & Akim Zedadka & 2 & Clermont & 3330 & 5.281 \\
70 & Presnel Kimpembe & 6 & Paris Saint-Germain & 1840 & 5.230 \\
71 & Ismail Jakobs & 3 & Monaco & 650 & 5.199 \\
\midrule
72 & Fábio & 3 & Nantes & 726 & 5.154 \\
73 & Thilo Kehrer & 2 & Paris Saint-Germain & 632 & 5.144 \\
74 & Cengiz Ünder & 10 & Marseille & 1047 & 5.079 \\
\midrule
75 & Léo Dubois & 2 & Lyon & 1246 & 5.060 \\
76 & Mattéo Guendouzi & 7 & Marseille & 1350 & 4.995 \\
77 & Facundo Medina & 4 & Lens & 1329 & 4.953 \\
\midrule
78 & Adrien Thomasson & 7 & Strasbourg & 1853 & 4.921 \\
79 & Nayef Aguerd & 6 & Rennes & 1698 & 4.908 \\
80 & Abdu Conté & 3 & Troyes & 695 & 4.891 \\
\midrule
81 & Javairô Dilrosun & 9 & Bordeaux & 675 & 4.845 \\
82 & Hamari Traoré & 2 & Rennes & 1878 & 4.836 \\
83 & Przemyslaw Frankowski & 3 & Lens & 1191 & 4.778 \\
\midrule
84 & Wissam Ben Yedder & 9 & Monaco & 1625 & 4.685 \\
85 & Vincent Le Goff & 3 & Lorient & 1440 & 4.669 \\
86 & Valentin Rongier & 2 & Marseille & 899 & 4.668 \\
\midrule
87 & Angelo Fulgini & 9 & Angers & 751 & 4.661 \\
88 & William Saliba & 5 & Marseille & 1800 & 4.652 \\
89 & Boubacar Kamara & 4 & Marseille & 1497 & 4.636 \\
\midrule
90 & Nuno Mendes & 3 & Paris Saint-Germain & 1246 & 4.599 \\
91 & Jason Denayer & 6 & Lyon & 630 & 4.591 \\
92 & Baptiste Santamaría & 8 & Rennes & 675 & 4.536 \\
\midrule
93 & Jonathan Gradit & 6 & Lens & 1710 & 4.535 \\
94 & Youssouf Fofana & 8 & Monaco & 1116 & 4.517 \\
95 & Florian Tardieu & 8 & Troyes & 1530 & 4.496 \\
\midrule
96 & Jordan Lotomba & 2 & Nice & 1410 & 4.454 \\
97 & Mihailo Ristic & 3 & Montpellier & 1150 & 4.439 \\
98 & Warmed Omari & 5 & Rennes & 1710 & 4.407 \\
\midrule
99 & Melvin Bard & 3 & Nice & 2470 & 4.350 \\
100 & Seko Fofana & 8 & Lens & 1861 & 4.345 \\
\bottomrule
\bottomrule
\end{tabular}

}
    \end{minipage}
    \vspace{0.4em}
    \caption{Ranking of Ligue 1 players in terms of GoT$^{(ind)}_{90}$.}\label{tab:fullrankingtotal}
\end{table}

\begin{table}[!ht]
    \centering
    \resizebox{0.4\columnwidth}{!}{
\begin{tabular}{ll}
    \toprule
    \toprule
        \textbf{Team} & \textbf{Formation cluster}  \\ 
        \midrule
        Angers & 3 \\
        Bordeaux & 3 \\
        Brest & 2 \\
\midrule
Clermont & 1 \\
Lens & 3 \\
Lille & 2 \\
\midrule
Lorient & 3 \\
Lyon & 1 \\
Marseille & 1 \\
\midrule
Metz & 3 \\
Monaco & 1 \\
Montpellier & 1 \\
\midrule
Nantes & 1 \\
Nice & 2 \\
Paris Saint-Germain & 1 \\
\midrule
Reims & 3 \\
Rennes & 1 \\
St Etienne & 4 \\
\midrule
Strasbourg & 3 \\
Troyes & 4 \\
        \bottomrule
        \bottomrule
    \end{tabular}}
\vspace{0.3em}
       \caption{The main formation clusters for each team in Ligue 1 in the 2021-2022 season.}
    \label{tab:formationclusters}
\end{table}

\end{document}